\documentclass[10pt,twocolumn,letterpaper]{article}
\PassOptionsToPackage{table,dvipsnames}{xcolor}
\usepackage{algorithm}
\usepackage{algpseudocode}

\usepackage{multirow}
\usepackage{graphicx}
\usepackage{amssymb}
\usepackage{adjustbox}
\usepackage{xcolor}
\makeatletter

\newcommand{\Rmnum}[1]{\expandafter\@slowromancap\romannumeral #1@}
\makeatother
\usepackage{iccv}  
\usepackage[accsupp]{axessibility}

\definecolor{iccvblue}{rgb}{0.21,0.49,0.74}
\newcommand{\blue}[1]{\textcolor{black}{#1}}
\newcommand{\justify}[1]{\textcolor{black}{#1}}
\usepackage[pagebackref,breaklinks,colorlinks,allcolors=iccvblue]{hyperref}
\usepackage{makecell}
\def\paperID{10585} 
\def\confName{ICCV}
\def\confYear{2025}

\title{Anti-Tamper Protection for Unauthorized Individual Image Generation}

\author{Zelin Li \quad Ruohan Zong \quad Yifan Liu \quad  Ruichen Yao \quad Yaokun Liu \quad Yang Zhang \quad Dong Wang\\
University of Illinois Urbana-Champaign \\
{\tt\small \{zelin3, rzong2, yifan40, ryao8, yaokunl2, yzhangnd, dwang24\}@illinois.edu}
}

\begin{document}
\maketitle
\begin{abstract}
{With the advancement of personalized image generation technologies, concerns about forgery attacks that infringe on portrait rights and privacy are growing. To address these concerns, protection perturbation algorithms have been developed to disrupt forgery generation. However, the protection algorithms would become ineffective when forgery attackers apply purification techniques to bypass the protection. To address this issue, we present a novel approach, \textbf{Anti-Tamper Perturbation (ATP)}. 
ATP introduces a tamper-proof mechanism within the perturbation. It consists of \textit{protection} and \textit{authorization} perturbations, where the protection perturbation defends against forgery attacks, while the authorization perturbation detects purification-based tampering. Both protection and authorization perturbations are applied in the frequency domain under the guidance of a mask, ensuring that the protection perturbation does not disrupt the authorization perturbation. This design also enables the authorization perturbation to be distributed across all image pixels, preserving its sensitivity to purification-based tampering.
ATP demonstrates its effectiveness in defending forgery attacks across various attack settings through extensive experiments, providing a robust solution for protecting individuals' portrait rights and privacy. Our code is available at: \href{https://github.com/Seeyn/Anti-Tamper-Perturbation}{https://github.com/Seeyn/Anti-Tamper-Perturbation} .}
\end{abstract}    
\section{Introduction}
\label{sec:intro}

\begin{figure*}[ht]
    \centering
    \includegraphics[width=0.95\textwidth]{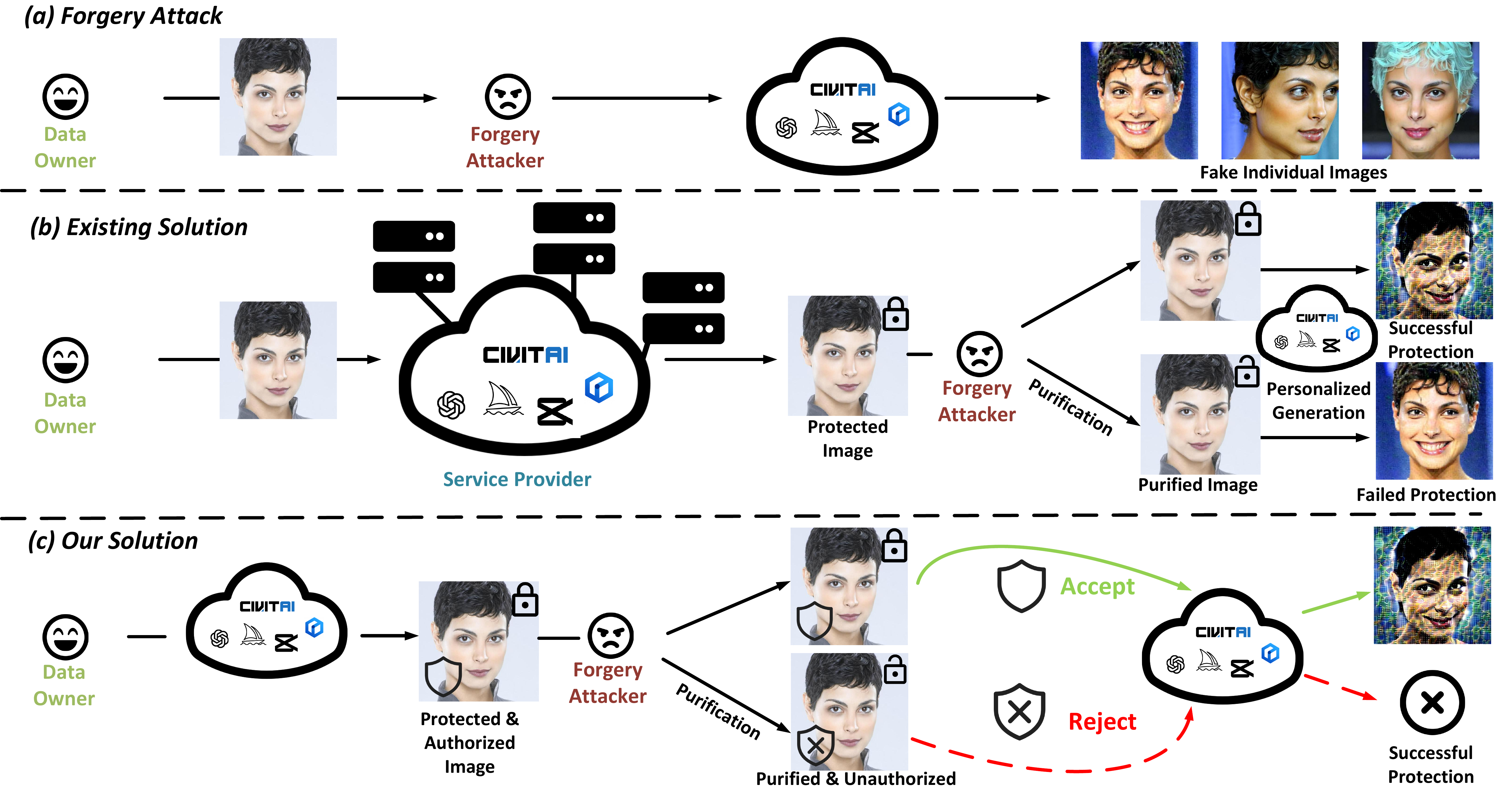}  
    \caption{
    (a) The forgery attacker generates fake individual images of the data owner by taking pictures from social media and submitting them to the service provider. 
(b) The data owner can inject protection perturbation into their images with assistance from the service providers, causing low-quality results if an attacker tries to generate fake individual images. However, the protection can fail if the attacker purifies the protected images. (c) Our solution: We propose Anti-Tamper Perturbation with a tamper-proof mechanism. If no purification is applied, the perturbation protects the data owner’s portrait rights and privacy by degrading the quality of generated images. Conversely, if a forgery attacker applies purification, the image becomes unauthorized, and the service provider rejects the generation request.}
    
    \label{fig1}
\end{figure*}

In recent years, with the development of personalized generation technology, online services for creating customized individual images have become widely available~\cite{TEXTINVERSE,LORA,DB}.
Users can easily create customized individual images with text prompts by submitting requests to online service providers such as Civitai\footnote{\href{https://civitai.com}{https://civitai.com}} and Midjourney\footnote{\href{https://www.midjourney.com}{https://www.midjourney.com}}. 
However, 
personalized generation technology also raises serious ethical and legal concerns. As shown in Figure~\ref{fig1}(a), forgery attackers can create fake individual images using online personalized generation services, infringing on the data owner’s portrait rights and privacy. In this scenario, the online service providers may inadvertently become accomplices of attackers by providing easily accessible services~\cite{Edwards2022}.


To defend the forgery attack, a set of protection methods~\cite{ANTIDB,ADVDM,METACLOAK,CAAT,INMARK}
have been proposed. They disrupt the personalized generation process by injecting protection perturbations into data owners' images. As shown in Figure~\ref{fig1}(b), the service provider injects protection perturbations to degrade the quality of generated images to prevent the forgery attacks. However, the protection perturbations can be easily purified by even naive purification methods (e.g., resizing or JPEG compression)~\cite{GRIDPURE,DBLP:journals/corr/abs-2406-12027}. The forgery attacker can purify the protection perturbations to bypass the protection schemes and generate fake individual images again.



\blue{     
As the essence of purification is to tamper the protection perturbation, the service provider can defend it by adopting a tamper-proof mechanism.  As shown in Figure~\ref{fig1}(c), the tamper-proof mechanism alerts the service provider when the protection perturbation is altered, allowing the service provider to counteract attacks by refusing generation requests from tampered images. This approach is similar to the Not Safe For Work (NSFW) content filtering mechanism~\cite{artsmart2024nsfw}, where the service provider inspects the generation request and its output to ensure no harmful content is present. Likewise, the tamper-proof mechanism detects and prevents attacks targeting tampering protection perturbations, enhancing the service provider’s defense capabilities.}
\justify{It is noteworthy that the tamper-proof mechanism is designed from the perspective of the service provider, where the purpose is to prevent the forgery attacker from misusing the service to launch a forgery attack.  The attackers may still be able to launch the attack on their own devices.  However, it is not the focus of this paper because it is not the service provider's responsibility and requires high-performance computing resources from the attacker rather than the effortless API calls used to exploit the service.}

\blue{Implementing a tamper-proof mechanism is challenging, as tamper-proofing requires protection perturbations to contain verifiable information that can be checked for potential tampering. However, protection perturbations are fundamentally adversarial noise, inherently designed to mislead deep learning models rather than encode structured information~\cite{DBLP:journals/corr/GoodfellowSS14}. Its creation depends on specific model architectures, loss functions, and input images, making it unsuitable for information embedding.} 

\blue{Considering this challenge, we propose a new perturbation design, \textbf{Anti-Tamper Perturbation (ATP)}. The ATP consists of two components:    \textit{protection} and \textit{authorization} perturbations. The protection perturbation is responsible for safeguarding the image against the forgery attacks. 
The authorization perturbation encodes an authorization message into the image. When purification occurs, the integrity of this message is disrupted, signaling that an unauthorized tamper attempt has occurred. 
It functions like a watermark~\cite{zhang2024editguard,FaceSigns}, yet the difference is that the existing watermark design is not sensitive to purification, making it incompatible with the authorization perturbation design.
When both protection and authorization perturbations function simultaneously, we can achieve the objective of tamper-proofing.
However, a fundamental challenge arises as the protection perturbation alters image information, making it conceptually a tampering manner. This creates an apparent dilemma: \textbf{\textit{the authorization perturbation must remain intact despite changes induced by protection perturbation while still being vulnerable to removal by purification-based tampering attacks}}. To address this challenge, we first adopt a Block Discrete Fourier Transformation (BDCT) to transform the image to the frequency domain.  We design a gradient descent algorithm to generate protection perturbations in the frequency domain. An authorization perturbation network is proposed to generate the authorization perturbation, embedding an authorization message in the frequency domain. Both perturbations can be guided by a binary mask, which specifies the regions in the frequency domain where perturbations should be applied. 
The mask ensures that the authorization perturbation remains intact even after the protection perturbation is applied, as they are positioned in different regions of the frequency domain determined by the mask.
Block Inverse Fourier Transformation (BIDCT) is then adopted to transform the image back to the pixel domain. Due to the transformation, the authorization perturbation is distributed across all pixels of the image, guaranteeing its sensitivity to purification-based tampering.}
Since ATP combines both protection and authorization perturbations, it is notable that ATP can work with various existing protection perturbation algorithms. The contributions of this work can be summarized as follows:



\begin{enumerate}
    \item To the best of our knowledge, our work is the first to introduce a tamper-proof mechanism for individual image generation protection, \blue{creating a novel approach to defend against forgery attacks with purification. }
    \item We design the \textbf{Anti-Tamper Perturbation (ATP)} to implement the tamper-proof mechanism. ATP comprises \textit{protection} and \textit{authorization} perturbations. \blue{The protection perturbation defends the image against forgery attacks, while the authorization perturbation remains unaffected by the protection perturbation yet retains sensitivity to purification-based tampering.}
    \item We evaluate the effectiveness of ATP in various attack scenarios through extensive experiments. The results show that ATP can be integrated with different protection perturbation designs.
    Existing solutions face an inevitable performance drop under attacks with purification. In contrast, ATP achieves a 100\% protection success rate due to the sensitive tamper-proof mechanism triggered by purification tampering.
\end{enumerate}



\section{Related Works}
\label{sec:related}

\paragraph{Individual Image Generation.}
The diffusion model is a leading technique in image generation~\cite{DBLP:conf/cvpr/RombachBLEO22,DBLP:conf/nips/DhariwalN21,DBLP:conf/nips/HoJA20}. It can generate an image by using text as a condition to guide the generation process~\cite{DBLP:conf/nips/SahariaCSLWDGLA22}. 
However, text conveys less detail than images, making it difficult to achieve specific results through text prompts alone, particularly for generating personalized content such as customized selfies~\cite{DB}.
To address this limitation, individual image generation methods (e.g., Text Inversion~\cite{TEXTINVERSE}, DreamBooth~\cite{DB}) were developed.
These approaches aim to ``learn" a unique token (e.g., \textit{sks}) that can represent a specific person or object. Diffusion models can apply this token to generate images with specific subjects~\cite{TEXTINVERSE, DB}.
Online service providers use individual image generation methods to provide the customized generation service, but the service might be misused for forgery attacks. ATP is designed to address this by preventing unauthorized use of personal images.
\paragraph{Protection Perturbation.}{Protection perturbation can be embedded within the image to safeguard against unauthorized generation.}
{The perturbation is typically generated by maximizing the diffusion model’s loss function, as first proposed by \citet{ADVDM}, who demonstrated that these perturbed images could act as adversarial examples for diffusion models.}~\citet{ANTIDB}  then introduced Anti-DB that enhanced AdvDM by incorporating Projected Gradient Descent (PGD) along with an Alternating Surrogate and Perturbation Learning strategy.
~\citet{CAAT}  presented CAAT to demonstrate that the cross-attention layer is critical in training diffusion models. This indicates that targeting the perturbation to disrupt image-text mapping can effectively enhance protection performance.
  {~\citet{METACLOAK}} proposed Metacloak that learns perturbations over a pool of surrogate models and applies the expectation-over-transformation technique to enhance the protection perturbation robustness against purification.

\paragraph{Perturbation Purification.}
 A key limitation of the protection perturbation is its protection performance drop when purification occurs. The purification can disrupt the perturbation’s integrity and weaken its protective capability~\cite{DBLP:journals/corr/abs-2406-12027, GRIDPURE}.
It is reported that naive purification techniques, such as image resizing and JPEG compression, would allow attackers to bypass the protection perturbation designs~\cite{DBLP:journals/corr/abs-2406-12027, GRIDPURE}. Furthermore,~\citet{GRIDPURE} introduced an advanced method, GridPure, to effectively purify protection perturbations. 
  As reported by~\cite{DBLP:journals/corr/abs-2406-12027,GRIDPURE}, the protection performance drop caused by purification remains a significant challenge. To address this issue, we introduce ATP, which shifts the focus from resisting purification to verifying perturbation integrity. ATP implements a tamper-proof mechanism for protection perturbation, allowing the service provider to reject generation requests of purified images.

\begin{figure*}[t]
    \centering
    \includegraphics[width=0.95\textwidth]{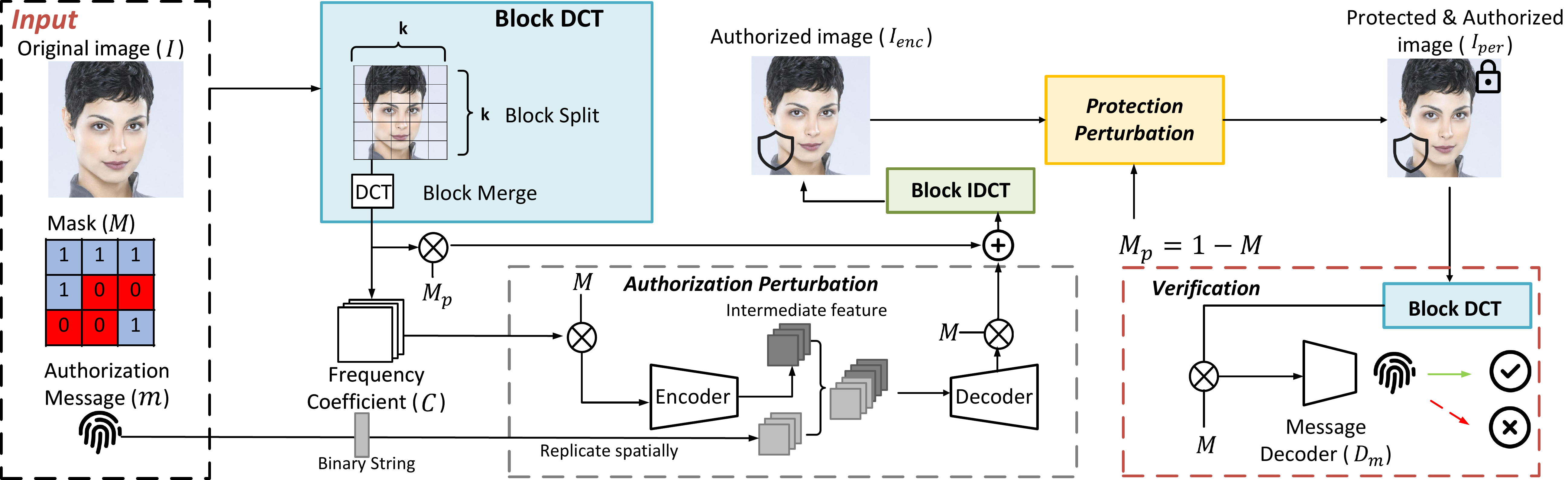}  
    \caption{Pipeline of the Anti-Tamper Perturbation. The original image is first transformed to the frequency domain using Block Discrete Cosine Transformation (BDCT). Guided by a binary mask, the authorization and protection perturbations are independently applied in the frequency domain to obtain a protected and authorized image.}
    \label{pipeline}
\end{figure*}
\section{Anti-Tamper Perturbation}

The pipeline of Anti-Tamper Perturbation is shown in Figure~\ref{pipeline}. The image is transformed by Block Discrete Cosine Transformation (BDCT) into the frequency domain. Guided by a binary mask, the authorization and protection perturbations are applied separately in the \textit{frequency domain} before being transformed back to the \textit{pixel domain}. 

\paragraph{Mask-Guided Perturbation Blending.}
 Both protection and authorization perturbation work by altering the image information. They can interfere with each other by altering the same image pixel. To address this problem, we propose distinguishing the perturbation by a mask as follows:
{\small
\begin{equation}
    P_{AP}(I) = M \odot P_{Auth}(I) + (1-M) \odot P_{Prot}(I),
\end{equation}}

\noindent where $I$ denotes the image, and $M$  represents the mask, composed of 0 and 1. 
The mask consists of the values $\{x|x\in\{0,1\}\}$ sampled from the Bernoulli distribution $x{\sim}\mathrm{Bernoulli}(p)$. As a result, we can adjust the hyper-parameter $p$ to control the ratio of 0s and 1s in the mask.
The $P_{AP}(\cdot)$, $P_{Auth}(\cdot)$, $P_{Prot}(\cdot)$ refer to the Anti-Tamper Perturbation, Authorization Perturbation, and Protection Perturbation, respectively. 
The mask allows the two perturbations to function separately, ensuring they do not interfere with each other. However, this design also introduces a limitation: the perturbations become distinguishable in the pixel space, which may allow purification methods to target the protection perturbation selectively.

To address this issue, we can adopt a transformation function that maps the image to the frequency domain, applying perturbations in the frequency domain: 
{\small
\begin{equation}
    P_{AP}(I) = F^{-1} [M \odot P_{Auth}(F(I)) + \\ 
    (1-M) \odot P_{Prot}\left(F\left(I\right)\right) ],
\end{equation}}

\noindent where $F(\cdot)$ and $F^{-1}(\cdot)$ denote the function projecting the image from the pixel domain to the frequency domain and its inverse function.  As each pixel value is a linear combination of frequency domain coefficients, both perturbations are uniformly distributed within the pixel domain. This approach ensures that the authorization and protection perturbations 
are indistinguishable in the pixel domain.  The uniform spread of the perturbation also improves the sensitivity of authorization perturbation to purification attempts.


\paragraph{Block Discrete Cosine Transformation.}
We employ a BDCT function as the transformation function $F$, which is more efficient than directly adopting Discrete Cosine Transform (DCT)~\cite{DCT} to the whole image~\cite{rao_yip_2014}. Specifically, we first divide image $I$ into non-overlapping blocks in the pixel domain. We record the position of the block in the pixel domain, for each block, we apply DCT as follows:

{\footnotesize
\begin{align}\label{eqa:DCT}
   &C_{u, v} = \alpha(u) \alpha(v) \sum_{i=0}^{N-1} \sum_{j=0}^{N-1}  I_{i, j} 
    \phi(u,i,N)
   \phi(v,j,N),
\end{align}}

\noindent where $\phi(u,i,N) = \cos \left( \frac{\pi (2u + 1) i}{2N} \right)$, $\alpha(i)=\sqrt{\frac{2}{N}}$ when $i=0$, $\alpha(i)=\sqrt{\frac{1}{N}}$ otherwise. $N$ represents the height and width of the block. $C$ denotes the frequency coefficients. Then, we merge the blocks based on their positions in the pixel domain. As a result, we can obtain a tensor of frequency coefficients with the same scale as the original image. We do the mask-guided authorization and protection perturbation to the tensor. To transform the image back from the frequency domain to the image domain, we apply the Block Inverse Discrete Cosine Transformation (BIDCT). We split the tensor into blocks in the same way as BDCT and apply Inverse-DCT to obtain the perturbed image in the pixel domain: {\begin{align}
   I_{i, j} = \sum_{u=0}^{N-1} \sum_{v=0}^{N-1} \alpha(u) \alpha(v) C_{u, v} \phi(u,i,N)\phi(v,j,N).
\end{align}}

\paragraph{Authorization Perturbation.}

Drawing inspiration from previous work on image steganography~\cite{hinet}, we consider embedding the authorization message in the frequency domain as a viable approach for authorization perturbation. 
Referring to the work~\cite{HIDDEN}, which has been proven to be effective in hiding information in the pixel domain, we use a convolutional autoencoder $f_{\theta}(\cdot)$ to complete the authorization perturbation. As illustrated in Figure~\ref{pipeline}, the encoder extracts intermediate feature maps from masked input frequency coefficients. The authorization message $m$, represented as a binary string of length $L$ composed of $\{0,1\}$, is spatially replicated and concatenated with the intermediate feature map. The decoder then reconstructs the coefficients from these modified feature maps. A message decoder $D_m$ is also trained to retrieve the encoded information from the reconstructed coefficients. The entire pipeline of the authorization perturbation can be formulated as follows:
{\small
\begin{equation}
    g_{\theta}(C) = (1-M)\odot C + M\odot f_{\theta}(M\odot C, m),
    \newline
    I_{enc} =  F^{-1}[g_{\theta}(F(I))],
\end{equation}}

\noindent where $I_{enc}$ denotes the image with authorization perturbation.
Referring to the model training of ~\cite {HIDDEN}, we incorporate an image reconstruction loss $\mathcal{L}_{rec}$ and an adversarial loss ${L}_{adv,G}$ to minimize alterations to the image content after perturbation.
For the accuracy of information hiding, we adopt a mask-guided message consistency loss and a regularization loss calculated in the frequency domain: 
{\small
\begin{equation}
    \mathcal{L}_{con} = ||D_m(M\odot f_{\theta}(C))  - m||^2_2, \mathcal{L}_{reg} = ||f_{\theta}(C)  - C||^2_2.
\end{equation}
}

\noindent The consistency loss facilitates message hiding, while the regularization suppresses significant changes in the frequency domain to keep the pixel domain values within the allowable range after inverse transformation.
The total loss is formulated as follows:
{\small
\begin{equation}
    \mathcal{L} =  \mathcal{L}_{con} + \lambda_{adv}\mathcal{L}_{adv,G} + \lambda_{rec}\mathcal{L}_{rec}  + \lambda_{reg}\mathcal{L}_{reg}.
\end{equation}}

\noindent By combining these loss functions, the authorization perturbation can embed messages covertly within the frequency domain of the image. 

\begin{algorithm}[!t]
\footnotesize
\caption{
Improved Frequency Domain PGD 
}
\begin{algorithmic}
\Require
Loss for perturbation $L$, Image for perturbation $I$, Guiding Mask $M_p$, Block DCT $F$, Block IDCT $F^{-1}$, PGD radius $\epsilon$, Step size $\alpha$.
\Ensure Perturbed Image $I_{per}$.
    \State $\nabla \leftarrow \frac{\partial \mathcal{L}}{\partial I}$
    \State $\nabla_{freq} \leftarrow M_p \odot F(\nabla)$
    \State $\nabla \leftarrow F^{-1}(\alpha\cdot \mathrm{sgn}(\nabla_{freq}))$
    \State $I_{per} \leftarrow I + \nabla$ 
    \State $I_{per} \leftarrow F^{-1}(\Pi_{\epsilon,F(I)}(F(I_{per})))$
    \Comment{$\Pi_{\epsilon,I}(\cdot)$ constrain its output within an $\epsilon$-ball around $F(I)$}
    \State \Return $I_{per}$
\end{algorithmic}
\label{algo:PGD2}
\end{algorithm}

\paragraph{Protection Perturbation.}
 {
All the protection perturbation algorithms essentially generate the perturbation according to gradients derived from the diffusion model loss function.
Based on the gradients, Projected Gradient Descent (PGD)~\cite{PGD} is the commonly used algorithm to update the perturbation.
As suggested by~\cite{LFAP}, in the frequency domain, performing mask-guided perturbation is analogous to omitting gradients that would alter frequency coefficients outside the target region for modification: 
{\small
\begin{equation}
    I_{per} \leftarrow \Pi_{\epsilon,I}(I + \alpha\cdot \mathrm{sgn}(F^{-1}(M_p\odot F(\nabla)))),
\label{FD_PGD}
\end{equation}}

\noindent where $\nabla$ denotes the gradient calculated from diffusion model loss and $\epsilon$ is the PGD radius. $\Pi_{\epsilon,I}(\cdot)$ constrains its output within an $\epsilon$-ball around $I$. $\mathrm{sgn}(\cdot)$ is the sign function and $\alpha$ is the step size. 
$M_p$ is a binary mask, where entries with value 1 indicate the frequency coefficients to be updated by the gradient, and entries with value 0 indicate those to remain unchanged. As shown in Figure~\ref{pipeline}, we can flip the mask $M$ for authorization perturbation to obtain this mask.}
 {
However, we find that updating the perturbation by Equation~\ref{FD_PGD} can invalidate the mask guidance. The reason is that each pixel value is a linear combination of all frequency coefficients. The modification of a single pixel value influences all the frequency coefficients.  The transformation of $\Pi(\cdot)$, $\mathrm{sgn}(\cdot)$ in the pixel domain inevitably alters the frequency coefficients the mask intended to preserve. To address this problem, we propose our improved frequency domain PGD algorithm in Algorithm~\ref{algo:PGD2}. $\Pi(\cdot)$, $\mathrm{sgn}(\cdot)$ are moved to apply in the frequency domain to ensure accurate modification of specific frequency coefficients based on the mask. \blue{A comparison experiment is placed in Appendix~\ref{sup2_1} to show the accuracy improvement.}}


Since existing protection perturbation methods rely on PGD for optimization, their algorithms can be directly adapted to use our Improved Frequency Domain PGD (Algorithm~\ref{algo:PGD2}), facilitating integration into ATP with minimal changes.


\section{Experiments}

In this section, we evaluate our ATP design from different perspectives. 
\blue{
First, we assess the effectiveness of ATP in defending against forgery attacks.  Three attack scenarios are evaluated: 1) attacks with purification, 2) attacks without purification, and 3) adaptive attacks for ATP. Second, we evaluate the aesthetic impact of ATP on the image. Third, we assess the robustness of the authorization perturbation to modifications induced by the protection perturbation and its sensitivity to purification tampering.
}

\paragraph{Datasets.} To train the authorization perturbation network, we utilize the FFHQ dataset~\cite{FFHQ}, which comprises 70,000 high-quality images of human faces. For generating the protection perturbation, we follow the dataset selection of \cite{ANTIDB}, using subsets of high-quality face image datasets CelebA-HQ~\cite{CelebA-HQ} and VGGFace2~\cite{VGGFace2}. We select 50 subjects from each dataset, each represented by eight images, and split them equally into two subsets. One subset is used for training the protection perturbation, and the other is reserved for the algorithm requirement of \cite{ANTIDB} and metric calculations.

\paragraph{Evaluation Metrics.}  For authorization performance evaluation, we calculate the bit error of the extracted and original authorization messages, denoted as \textbf{Bit-error}.     This metric represents how accurately the authorization perturbation hides the information in the image. The Bit-error increases when the purification attempts alter the authorization perturbation.
Following the choice of \cite{METACLOAK,ANTIDB}, for protection performance evaluation, we take Stable Diffusion v2-1~\cite{DBLP:conf/cvpr/RombachBLEO22} as the base generation model and apply DreamBooth~\cite{DB} for personalized image generation.
We generate 16 images for each subject
by prompt ``a photo of \textit{sks} person." (Additional results for a different generation model, personalized generation algorithm, and prompt can be found in Appendix~\ref{sup2_2} \& \ref{sup2_3}).
We assess the protection performance using four metrics:
1) \textbf{CLIP-IQAC}: CLIP-IQAC is proposed by \cite{METACLOAK} and adapted from CLIP-IQA \cite{CLIP-IQA}, this metric evaluates human images in quality.
2) \textbf{LIQE}: LIQE is an image quality assessment metric that aligns well with human perception and applies to human and non-human images \cite{LIQE}.
3) \textbf{Face Detection Failure Rate (FDFR)}: FDFR measures the failure rate of face detection using Retinaface \cite{retinaface}.
4) \textbf{Identity Score Matching (ISM)}: ISM evaluates the identity consistency between the generated image and the original image. The identity embedding of generated images and the original images are extracted by Arcface~\cite{Arcface}. 
ISM is computed by measuring the cosine similarity between the generated image's identity embedding and the average embedding of the original images.

The four metrics assess the impact of perturbations on generation quality from different perspectives. However, there is no clear standard for defining a ``successful protection."
As a result, we propose the Protection Success Rate (PSR), which quantifies the effectiveness of protection by establishing a threshold for CLIP-IQAC, where any generated image with a quality score below this threshold is deemed as successfully protected.
We also define a Bit-error threshold. Any generation request with an unauthorized image whose Bit-error exceeds the threshold will be rejected. 
 If a user submits four images for generation, and at least one of them is unauthorized, the service provider can reject the generation request. This is considered a successful protection, resulting in a PSR of 1.0. The experiments about why we select the CLIP-IQAC to calculate PSR and how we set the threshold for Bit-error and CLIP-IQAC are placed in Appendix~\ref{sup1_2} \& \ref{sup1_3}.

\paragraph{Baselines.} As ATP can be integrated with different protection algorithms, we select four state-of-the-art protection algorithms for our experiments: Anti-DB (2023)\cite{ANTIDB}, AdvDM (2023)\cite{ADVDM}, CAAT (2024)\cite{CAAT}, MetaCloak (2024)\cite{METACLOAK}. 
By modifying the gradient descent approach of these protection algorithms as described in Algorithm \ref{algo:PGD2}, we can adapt the baseline protection perturbation to ATP. \blue{The details of ATP's implementation are in Appendix~\ref{sup1_1}. The mask ratio $p=0.5$ and the BDCT block width $N=16$ are determined based on the ablation studies in Appendix~\ref{sup2_4} and \ref{sup2_5}.}

\paragraph{\blue{Protection Performance Under Attacks with Purification.}} \label{Protection Performance Under Attack Scenario}
\blue{In this experiment, we demonstrate the ATP design is effective 
    for protecting individual image generation under attacks with purification.}
We select two purification methods for naive purification: JPEG compression and image resizing, which have been reported to be effective \cite{GRIDPURE, METACLOAK} to purify the protection perturbation. For the advanced purification method, we choose GridPure (2024)~\cite{GRIDPURE}.
\begin{figure}[tbp]
    \begin{subfigure}[b]{\textwidth}
        \includegraphics[width=0.5\textwidth]{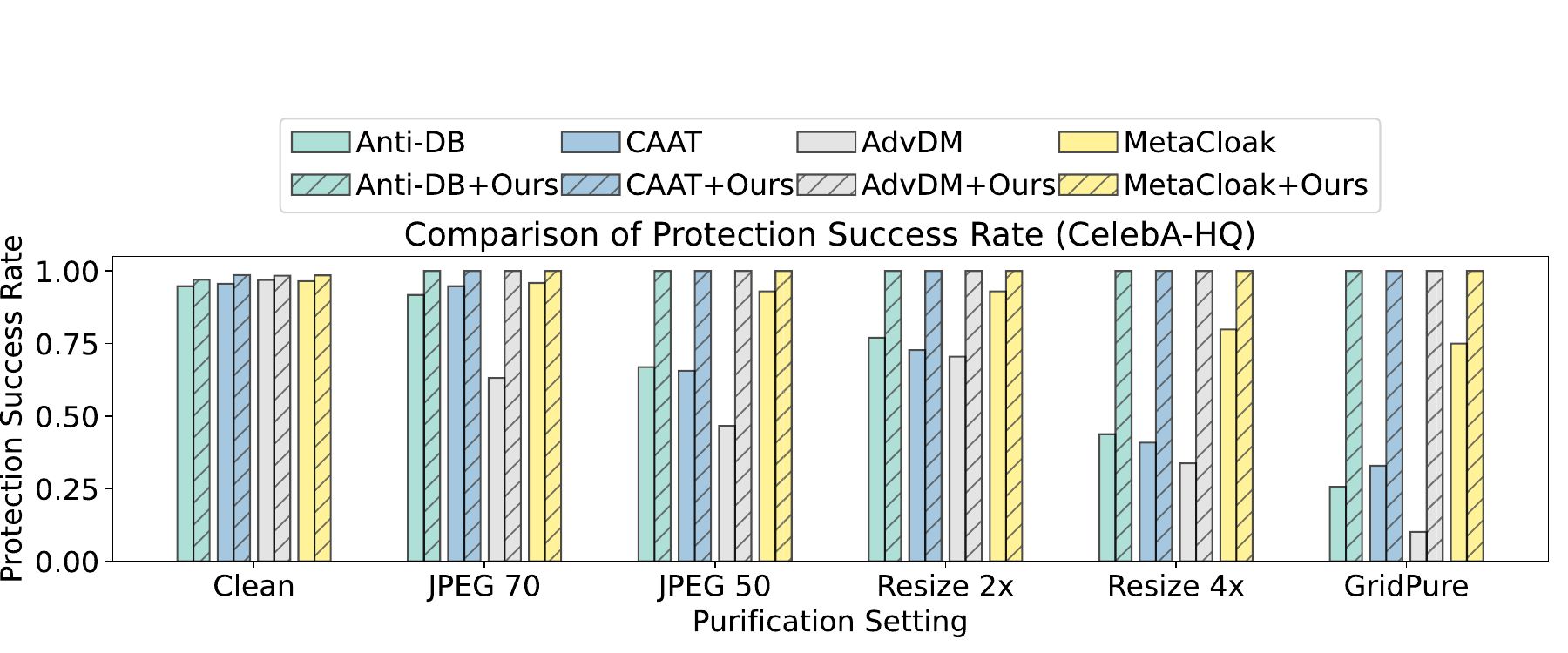}
    \end{subfigure}

    \vspace{-3mm}
    \begin{subfigure}[b]{\textwidth}
        \includegraphics[width=0.5\textwidth]{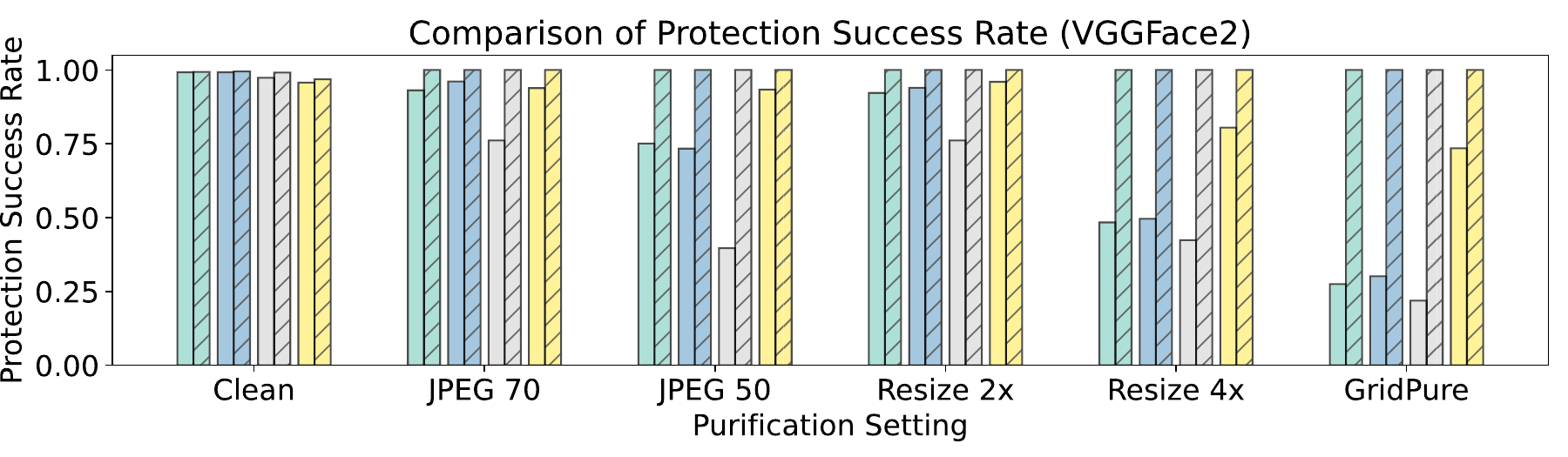}
    \end{subfigure}
     \vspace{-8mm}
    \caption{Comparison of Protection Success Rate for different methods across various purification settings. }
    \label{fig: attack scenario}
\end{figure}
The PSR metric is used to evaluate different methods across two datasets, and the results are shown in Figure~~\ref{fig: attack scenario}.  
The results show, in the ``clean" condition where no purification is applied, all methods achieve relatively high protection success rates. In this setting, ATP’s tamper-proof mechanism is not triggered, and no unauthorized image is found.
When purification is applied, the performance of different methods faces an unavoidable reduction in protection performance. Such performance decline is especially noticeable in settings like Resize 4x, JPEG 50, and GridPure (Visual results of the generated images after purification are provided in Appendix~\ref{sup2_9}). Even MetaCloak, a robust protection perturbation algorithm designed against purification, its PSR still inevitably drops. However, when combined with ATP, MetaCloak and other baselines all achieve a 100\% Protection Success Rate, as ATP’s tamper-proof mechanism reliably detects various purifications and triggers generation rejection. It indicates the effectiveness of ATP in defending against attacks with purification.

\begin{table}[t!]
\centering
\setlength{\tabcolsep}{0.5pt}
\renewcommand{\arraystretch}{1.2} 
\scalebox{0.75}{

\begin{tabular}{c|cccc|cccc}
\hline

\multirow{2}{*}{} & \multicolumn{4}{c|}{\textbf{CelebA-HQ}} & \multicolumn{4}{c}{\textbf{VGGFace2}} \\ \cline{2-9} 
                  & \multirow{2}{*}{\makecell{CLIP-\\IQAC $\downarrow$}} & \multirow{2}{*}{LIQE $\downarrow$} & \multirow{2}{*}{ISM $\downarrow$} & \multirow{2}{*}{FDFR $\uparrow$} & \multirow{2}{*}{\makecell{CLIP-\\IQAC $\downarrow$}} & \multirow{2}{*}{LIQE $\downarrow$} & \multirow{2}{*}{ISM $\downarrow$} & \multirow{2}{*}{FDFR $\uparrow$} \\ 
                  &                                       &                    &                    &                   &                      &                   &                   &                   \\ \hline
Anti-DB           & -0.2870	&1.1168	&0.4619	&0.4575	&-0.4340	&1.0282	&0.3201	&0.6788              \\ \hline
\rowcolor{gray!40}
Anti-DB+Ours   \rowcolors{green!70}   &-0.3139	&1.0741	&0.4647	&0.5213	&-0.3864	&1.0984	&0.2549	&0.8188             \\ \hline
AdvDM           &-0.3361	&1.0287	&0.4166	&0.6638	&-0.3797	&1.0163	&0.4061	&0.6100\\ \hline
\rowcolor{gray!40}
AdvDM+Ours       &-0.3621	&1.0312	&0.4119	&0.6675 &-0.4370	&1.0355	&0.2540	&0.8638             \\ \hline
CAAT            &-0.3261	&1.0872	&0.4577	&0.4700	&-0.5008	&1.0139	&0.2977	&0.7825   \\ \hline
\rowcolor{gray!40}
CAAT+Ours       &-0.3568	&1.0768	&0.4315	&0.6338	&-0.4673	&1.0291	&0.2497	&0.7863              \\ \hline
MetaCloak        &-0.3418	&1.4243	&0.4911	&0.4850	&-0.3358	&1.2247	&0.4857	&0.5025   \\ \hline
\rowcolor{gray!40}
MetaCloak+Ours   &-0.3639	 &1.3140	 &0.4048	 &0.7038	 &-0.3628	 &1.1573	 &0.4179	 &0.6338               \\ \hline
\end{tabular}
}
\caption{Quantitative results for CelebA-HQ and VGGFace2 datasets across various metrics. }
\label{tab: MR}
\end{table}

\paragraph{\blue{Protection Performance Under Attacks without Purification.} }

\begin{figure}[t]
    \centering
    \includegraphics[width=0.9\columnwidth]{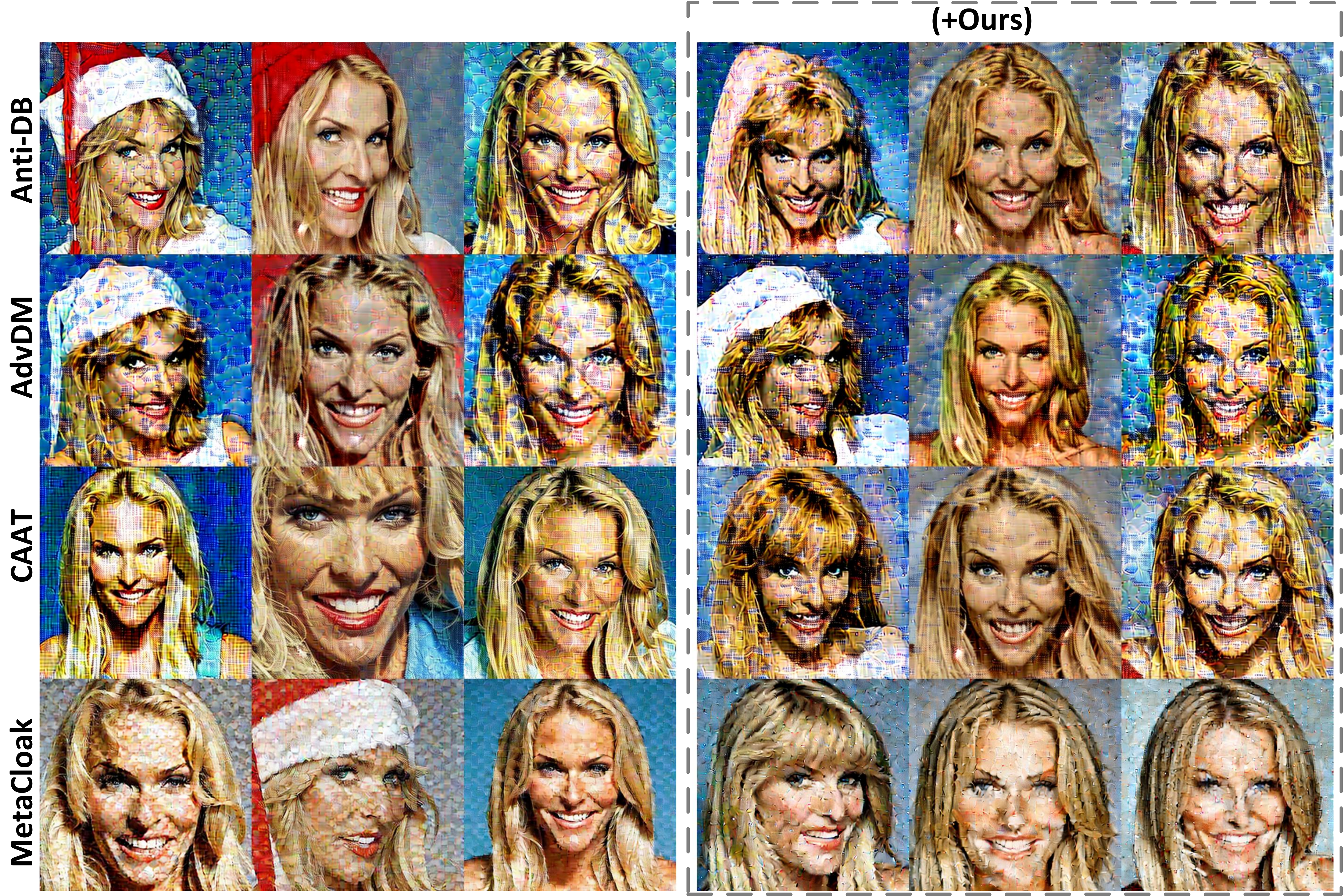}  
    \caption{Qualitative comparison of original perturbation algorithms and their ATP modified versions in CelebA-HQ. More visual results from the VGGFace2 are provided in Appendix~\ref{sup2_10}.}
    \label{fig: quantitative comparison}
\end{figure}

In this experiment, we show the protection performance of adopting an ATP design under attacks without purification, where the tamper-proof mechanism is not triggered.
The results are shown in Table~\ref{tab: MR}. 
By adding baseline protection perturbation with ATP design, we observe a consistent increase in FDFR, a decrease in ISM and CLIP-IQAC in most cases, and similar results for LIQE.
As illustrated in Figure~\ref{fig: quantitative comparison}, each baseline retains its original capability to degrade image quality after adapting to ATP. 
The results show that although tamper-proof mechanism will not be triggered under attacks without purification, ATP can still achieve performance comparable to the original protection perturbation.

\paragraph{\blue{Protection Performance Under Adaptive Attacks.}}

\blue{For the adaptive attack, we consider three settings:  
(1) the attacker knows the mask value but not the BDCT hyper-parameters,  
(2) the attacker knows the hyper-parameters but not the mask value, and  
(3) the attacker knows both.  
We conduct experiments on MetaCloak with ATP using the VGGFace2 dataset and apply a rounding function to purify perturbations in the frequency domain.  
In setting (1), we change the BDCT block size from 16 to 8, leading to a significant increase in Bit-error from \(6.25 \times e^{-4}\) to \(3.18 \times e^{-1}\), causing the authorization verification to fail.  
For (2), we apply the rounding action to all frequency coefficients, which raises the Bit-error from \(6.25 \times e^{-4}\) to \(5.04 \times e^{-1}\),  also resulting in verification failure.  
In (3), the attack successfully bypasses verification, reducing the Protection Success Rate to 0.33.  
While ATP is ineffective in setting (3), its strong resistance in (1) and (2) highlights its robustness against attackers with partial knowledge. 
The vulnerability in (3) is due to the complete leakage of both BDCT hyper-parameters and the mask value, which is extremely unlikely to happen. Even without considering variations in BDCT hyperparameters, the search space of a binary random mask (with a size of $512 \times 512 \times 3$) remains $C^{393216}_{786432}\approx2^{786414}$ given a known mask ratio of 0.5. Such a large search space makes it practically impossible for an attacker to retrieve the mask without human-induced leakage.
Therefore, ATP remains a viable defense under adaptive attack conditions, particularly when attackers have only partial knowledge.}

\begin{figure}[ht]
    \centering
    \includegraphics[width=0.9\linewidth]{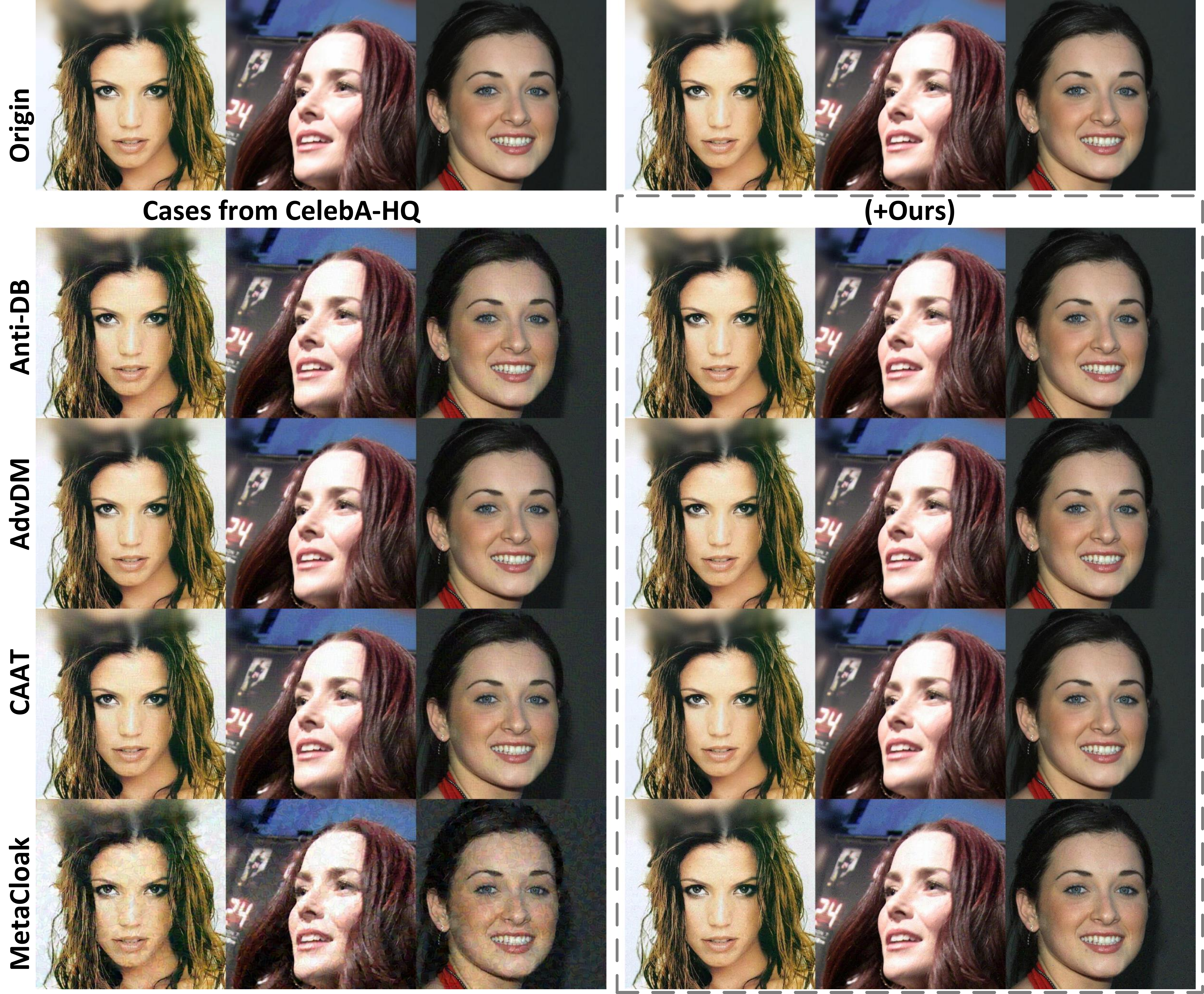}  
    \caption{\blue{Perturbed images of different methods from CelebA.}}
    \label{fig: perturbed imgs}
\end{figure}
\begin{table}[h]
\centering
\scriptsize
\renewcommand{\arraystretch}{0.8}

\begin{adjustbox}{width=\linewidth}
\scriptsize
\begin{tabular}{lcc|cc}
\toprule
\multirow{2}{*}{} & \multicolumn{2}{c|}{\textbf{CelebA}} & \multicolumn{2}{c}{\textbf{VGGFace}} \\
& CLIP-IQAC$\uparrow$ & ISM$\uparrow$ & CLIP-IQAC$\uparrow$ & ISM$\uparrow$ \\
\midrule
Anti-DB & 0.3925	&0.7569	&0.4220	&0.7224 \\
\rowcolor{gray!40} Anti-DB+Ours & 0.5744	&0.7716	&0.4195	&0.6934 \\
AdvDM & 0.5115	&0.7686	&0.4986	&0.7289 \\
\rowcolor{gray!40}AdvDM+Ours & 0.5785	&0.7703	&0.4280	&0.6910 \\
CAAT & 0.3886	&0.7608	&0.4211	&0.7230 \\
\rowcolor{gray!40}CAAT+Ours & 0.5733	&0.7712	&0.4519	&0.7088 \\
MetaCloak & -0.0139	&0.7126	&0.2046	&0.6766 \\
\rowcolor{gray!40}MetaCloak+Ours & 0.4923	&0.7581	&0.4390	&0.6982 \\
\bottomrule
\end{tabular}

\end{adjustbox}

\caption{\blue{Quantitative evaluation of the aesthetic impact of different perturbation algorithms across two datasets.}}
\label{asethetic}
\end{table}
\paragraph{\blue{Aesthetic Impact of Perturbation.}}
The perturbation may degrade image quality and compromise identity consistency, potentially leading to poor aesthetics and discouraging image owners from adopting such techniques. In this experiment, we evaluate the aesthetic impact of ATP and other protection perturbations using ISM and CLIP-IQAC. Since both metrics are computed on the perturbed images rather than the generated ones, higher values indicate less aesthetic impact.
As shown in Table~\ref{asethetic}, the combination with ATP generally leads to increases in both ISM and CLIP-IQAC, suggesting that ATP introduces minimal aesthetic degradation. This observation is further supported by the qualitative comparisons in Figure~\ref{fig: perturbed imgs}. Additional visual results from VGGFace2 are provided in Appendix~\ref{sup2_8}.





\begin{table}[!tbp]
    \centering
    \tiny
    \renewcommand{\arraystretch}{1} 
    \resizebox{\linewidth}{!}{
    \tiny
    \begin{tabular}{c|ccc|c}
        \toprule[1 pt]
        & BDCT & Improved-PGD & Mask & Bit-error ($e^{-3}$)$\downarrow$  \\
        
        \hline
        (a) & & & & 349.84  \\
        \hline
        (b) & & & $\checkmark$ & 42.031  \\
        \hline
        (c) & $\checkmark$ & & & 360.31  \\
        \hline
        (d) & $\checkmark$ & & $\checkmark$ & 81.719  \\
        \hline
        Ours & $\checkmark$ & $\checkmark$ & $\checkmark$ & 0.4688  \\
        \bottomrule[1 pt]
    \end{tabular}}
    \caption{Comparison of different fusion designs with Bit-error values.  \textbf{BDCT} determines whether the BDCT is used to transform the image to the frequency domain; If not, the perturbation is applied directly in the pixel domain. \textbf{Improved-PGD} indicates whether we adopt Algorithm~\ref{algo:PGD2}; If not, we adopt Equation~\ref{FD_PGD}. \textbf{Mask} specifies whether we use the guiding mask of ratio 0.5; If not, the perturbations are applied without the guidance of the mask.}
    \label{tab: MC_2}
\end{table}

\paragraph{\blue{Robustness to Protection Perturbation.}}

\label{sec:perturbation fusion}
\begin{figure}[tbp]
    \centering
    \includegraphics[width=0.9\columnwidth]{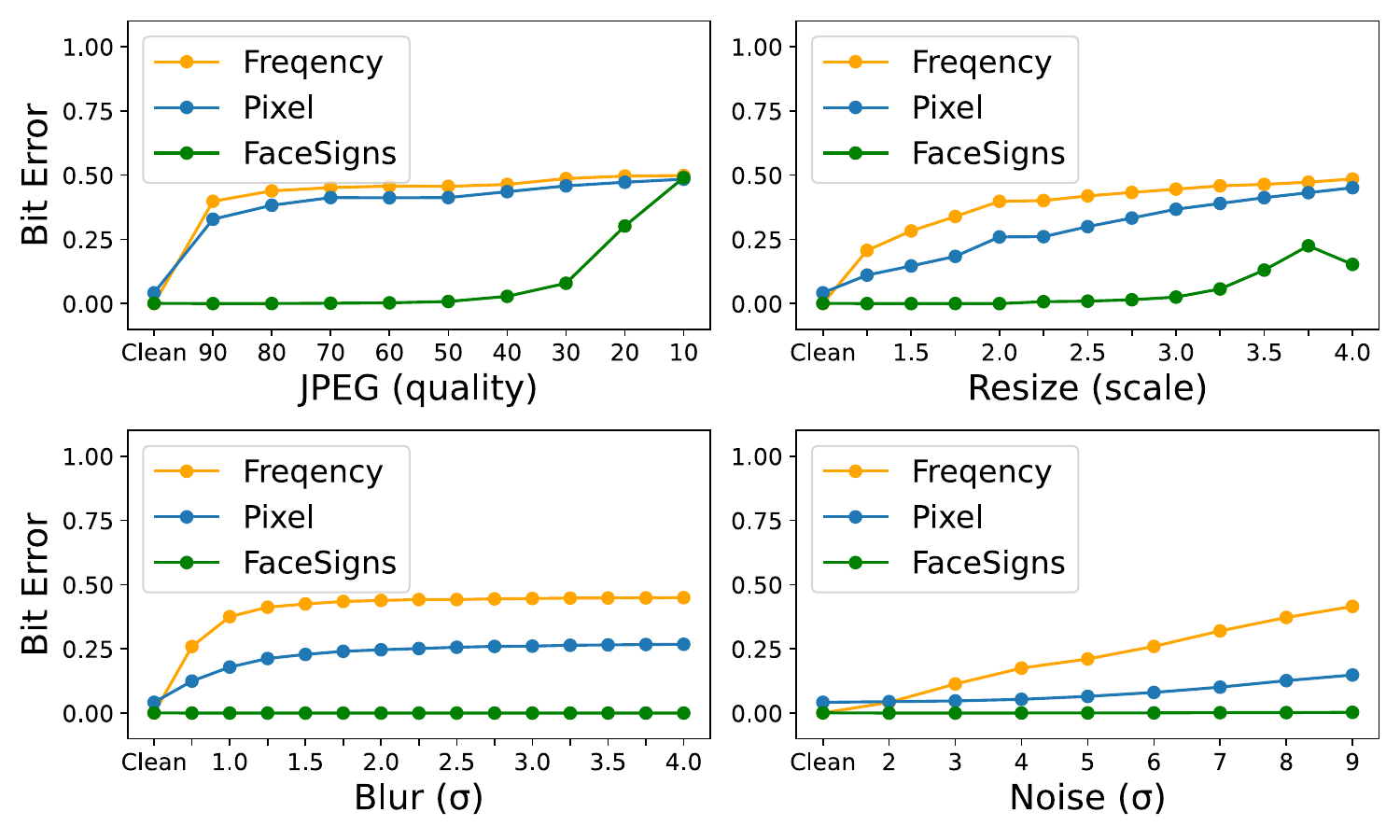}  
    \caption{Sensitivity of ATP to different types of purification. 
    The x-axis indicates the hyperparameter of different purifications, while the y-axis indicates the Bit-error.}
    \label{fig:Ab_pixelfreq}
\end{figure}
In this experiment, we evaluate different perturbation fusion strategies to analyze the factors that influence the robustness of the authorization perturbation against interference from the protection perturbation.
We incorporate Anti-DB into ATP for this experiment on CelebA-HQ. VGGFace2 results are provided in Appendix~\ref{sup2_7}. The robustness is evaluated by the Bit-error, the lower the error, the higher the robustness.  
From the results in Table \ref{tab: MC_2}, we observe that settings (a) and (c) exhibit a large Bit-error, underscoring the importance of mask guidance. Comparing setting (b) with ``Ours", we find that encoding messages in the frequency domain leads to less Bit-error than in the pixel domain. The comparison between setting (d) and ``Ours" indicates that 
Equation~\ref{FD_PGD}
invalidates the mask guidance, which results in an increased Bit-error. The experiment highlights the importance of using a mask to guide the fusion of authorization and protection perturbations. It also demonstrates that frequency-domain authorization perturbation is more effective at concealing information than its pixel-domain counterpart. Moreover, the results underscore the necessity of Algorithm~\ref{algo:PGD2} for accurate mask guidance.


\paragraph{Sensitivity to Purification.} 

In this experiment, we compare the sensitivity of frequency-domain ATP versus pixel-domain ATP to purification.
\blue{We also compare the performance with the existing watermark method FaceSigns \cite{FaceSigns}.}
For the pixel-domain design, we adopt setting (b) from the robustness experiment. 
We incorporate Anti-DB into ATP for this experiment on CelebA-HQ. VGGFace2 results are provided in Appendix~\ref{sup2_7}.
We select four types of purification to purify the images with ATP: Resize, JPEG, Gaussian Blur, and Gaussian Noise. The hyperparameters include the JPEG compression quality, downsampling scale for resizing, the sigma value for Gaussian blur (with a kernel size of 3), and Gaussian noise. 
The results are shown in Figure~\ref{fig:Ab_pixelfreq}. The Bit-error rises sharply as purification intensity increases, indicating the high sensitivity of ATP to various types of purification. Compared to its pixel-domain counterpart, the frequency-domain implementation of ATP achieves a lower Bit-error when no purification is applied, but exhibits a steeper increase under purification. This suggests that frequency-domain authorization perturbations are both more effective at concealing messages and more sensitive to purification.
This increased sensitivity arises from the fact that frequency-domain perturbations are uniformly distributed across the pixel domain, making them more vulnerable to purifications. Additional theoretical analysis of this phenomenon is provided in Appendix~\ref{sup1_4}. In contrast, FaceSigns exhibits low sensitivity to purification due to its robustness against such operations, indicating that existing solutions are not directly compatible with ATP design.


\section{Discussion}
While ATP demonstrates strong effectiveness in our evaluated scenarios, it introduces additional computational cost and degrades to a regular protection perturbation when attackers operate on their own devices to bypass the verification process. To address concerns about deployment cost, we provide a scalability analysis in Appendix~\ref{sup2_11}, showing that ATP remains computationally affordable for the service provider.
Despite these limitations, ATP still offers a practical and robust defense mechanism by preventing service providers from inadvertently contributing to forgery attacks. Meanwhile, since image generation requires significantly more computational resources than purification attacks, the need for a generation-capable device raises the barrier for potential attackers.
These limitations also point to a promising direction for future work: eliminating the need for an explicit verification process from the service provider. 
One possible direction is to design authorization perturbations whose disruption degrades the generation result, thereby removing the reliance on explicit verification. 


\section{Conclusion}


This paper introduces a novel perturbation design called Anti-Tamper Perturbation (ATP), motivated by the challenge that forgery attackers can bypass protection perturbation defenses through purification techniques. To address this issue, the ATP incorporates a tamper-proof mechanism. When purification occurs, the integrity of the ATP is compromised, signaling to the service provider that the image has been altered. This allows the provider to reject the unauthorized image and mitigate the threat of forgery attacks with purification.  Extensive experiments conducted on two datasets demonstrate that ATP consistently outperforms state-of-the-art baselines in preventing forgery attacks under a wide range of purification methods. This paper highlights the potential of ATP as a solution for resisting forgery attacks, offering greater feasibility for safeguarding portrait rights and personal privacy in real-world scenarios. 


{
    \small
    \bibliographystyle{ieeenat_fullname}
    \bibliography{main}
}
\appendix
\setcounter{page}{1}
\maketitlesupplementary

In the appendix, we will include more experimental results and the detailed settings for anti-tamper perturbation (ATP).

\section{Details of the Anti-tamper Perturbation}
\subsection{Hyper-parameter configuration}\label{sup1_1}
\paragraph{Experiment Environment.}

All experiments were conducted on a server equipped with 4 L40S GPUs (each with 48G) and an Intel(R) Xeon(R) Gold 6426Y CPU. The system had 251 GB of RAM. The software environment included Pytorch 2.4.1  running on Ubuntu 22.04.4 LTS, with CUDA 12.3 and cuDNN 9.1.0.70  for GPU acceleration. We didn't do distributed training, so the experiment can be conducted using one GPU.

\paragraph{Authorization Perturbation Hyper-Parameters.}
The authorization perturbation network is trained on FFHQ for 65,000 steps with a batch size of 8.
For the weights of the loss function: $\lambda_{adv}=1e-3$, $\lambda_{rec}=0.7  $, $\lambda_{reg}=10$. The length of the authorization message $m$ is 32, and the default mask ratio $p$ is 0.5.

\paragraph{Protection Perturbation Hyper-Parameters.}
APT can adopt the existing protection design, and different baselines have varying choices for PGD radius and step size.  Unlike the baselines, our method performs calculations in the frequency domain, so we did not select the same hyperparameters as the baseline.

\begin{table}[ht]
\centering
\begin{tabular}{@{}lcc@{}}
\toprule[1.5pt]
\textbf{Method}       & {\textbf{CelebA-HQ}} & \textbf{VGGFace2} \\ \cmidrule(lr){2-2} \cmidrule(lr){3-3}
                      & \textbf{Radius / Step Size}   & \textbf{Radius / Step Size} \\ \midrule
Anti-DB+ours           & 5e-2 / 5e-3                 & 250e-3 / 25e-3             \\
AdvDM+ours            & 1e-1 / 2e-3                 & 250e-3 / 25e-3             \\
CAAT+ours             & 5e-2 / 5e-3                 & 250e-3 / 25e-3             \\
MetaCloak+ours        & 150e-3 / 5e-3                 & 200e-3 / 5e-3              \\ \bottomrule[1.5pt]
\end{tabular}
\caption{PGD Radius and Step Size for different methods on CelebA-HQ and VGGFace2.}
\label{tab:comparison}
\end{table}
\noindent We observed the loss performance after adapting the baseline to our algorithm and selecting the appropriate PGD radius and step sizes. However, we did not perform detailed hyperparameter tuning experiments, as our main objective was to demonstrate that our method does not degrade the baseline's protection performance.
\subsection{Metric Selection}\label{sup1_2}

\begin{figure}[!t]
    \centering
    \includegraphics[width=0.9\columnwidth]{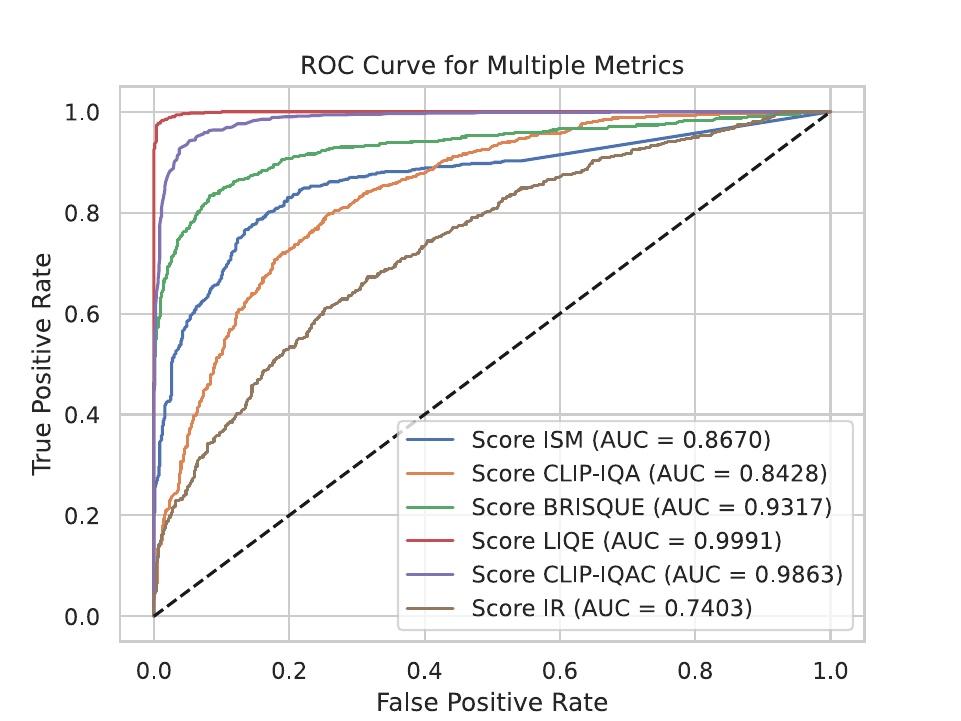}  
    \caption{The ROC curve of different metrics.}
    \label{fig: roc}
\end{figure}

To select suitable metrics for evaluating the protection perturbation, we choose six metrics from the metrics adopted by existing works~\cite{ANTIDB,CAAT,METACLOAK}:
ISM~\cite{ANTIDB}, CLIP-IQA~\cite{CLIP-IQA}, BRISQUE~\cite{BRISQUE}, LIQE~\cite{LIQE}, CLIP-IQAC~\cite{METACLOAK}, IR~\cite{IR}. 
When the model can't detect the face, the ISM value is set to -1 to guarantee that all generated images can get a corresponding ISM value.
We first generate individual images using unprotected images from CelebA-HQ and those protected with Anti-DB. For each subject, we generate 16 images (50 subjects in total). We then calculate the value of the generated image's six metrics accordingly. We assume that the Anti-DB can often successfully protect the image when attackers do not make purification attempts. As a result, if the metric can classify images generated from protected images and those generated from unprotected images, it should be a reliable metric for evaluating protection performance. 
Images generated from protected images are categorized as negative samples, while those generated from unprotected images are categorized as positive samples. We then draw the ROC curves of the protection performance metrics, as shown in Figure~\ref{fig: roc}.
\begin{figure*}[t]
    \centering
    \begin{subfigure}[b]{0.48\textwidth}
        \centering
        \includegraphics[width=\textwidth]{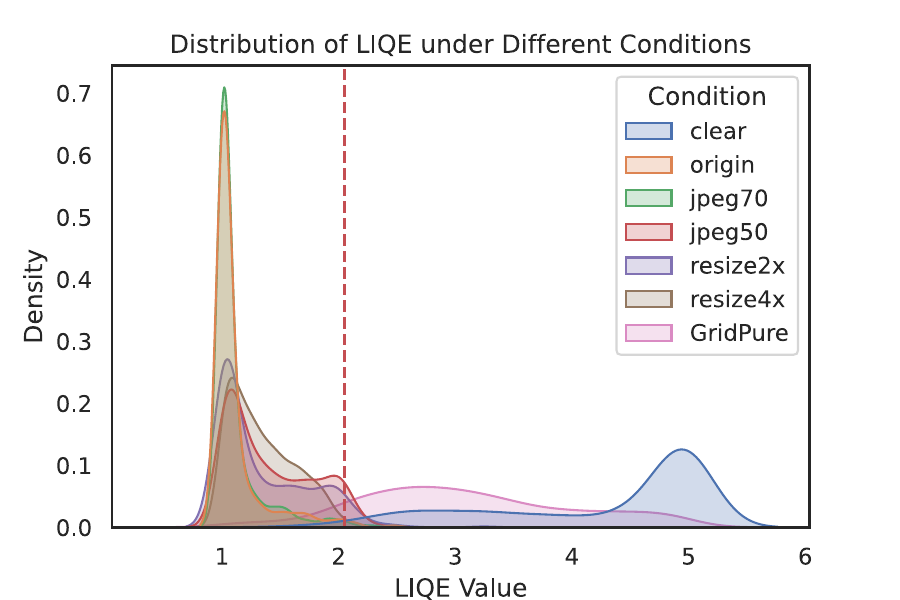}  
        \caption{}
        \label{fig:liqe_comparison}
    \end{subfigure}
    \hspace{0.5mm}
    \begin{subfigure}[b]{0.48\textwidth}
        \centering
        \includegraphics[width=\textwidth]{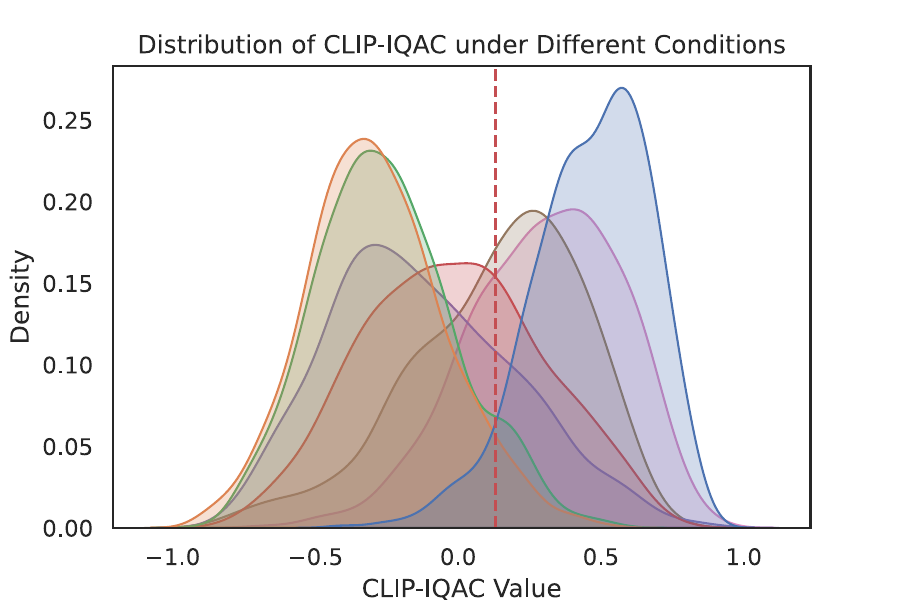}  
        \caption{}
        \label{fig:clip_iqac_comparison}
    \end{subfigure}
    \caption{(a) Distributions of generated images evaluated by LIQE metric. (b) Distributions of generated images evaluated by CLIP-IQAC metric. The red dashed line illustrates the PSR threshold.} 
    \label{fig: roc2}
\end{figure*}
Among the metrics, CLIP-IQAC and LIQE show the highest AUC values, demonstrating the strongest discriminatory ability. 
As a result, we adopt them for the \textbf{Standard Protection Performance Comparison} in Section 4 (FDFR and ISM are also adopted, as ISM is the only metric among them that is directly related to facial identity. Furthermore, FDFR and ISM are typically computed together \cite{ANTIDB}.). For the experiment of the \textbf{Protection Performance Under Attack Scenario}, we need to select one metric for calculating the Protection Success Rate (PSR). We use the property of the ROC curve to decide the threshold of PSR. We select the threshold that can minimize $\sqrt{(1-TPR)^2+FPR^2}$, where $TPR$ denotes true positive rate and $FPR$ denotes false positive rate.
\textbf{The threshold for CLIP-IQAC and LIQE are 0.1318359375, 2.05078125, respectively}.

Subsequently, we evaluate the performance of these metrics in capturing the impact of purification attempts on the protection mechanism. The distribution of the metrics for generated images is visualized through kernel density estimation. Specifically, ``clear" and ``origin" represent the generation results using unprotected and protected images, respectively. At the same time, the remaining categories correspond to the outcomes of applying the respective purification methods to protected images before generation.

As shown in Figure~\ref{fig: roc2}, 
 the results demonstrate that CLIP-IQAC and LIQE effectively reflect the influence of purification attempts. Notably, following ``resize 4x'', ``jpeg 50'', and ``GridPUre'', the resulting distributions exhibit a convergence trend toward those of  ``clear."  However, it can be observed that the PSR threshold of LIQE fails to capture the trend, as the majority of the samples fall to the left of the threshold. In contrast, CLIP-IQAC does not exhibit this issue, making it the preferred choice for calculating PSR.

\subsection{Threshold Setting}\label{sup1_3}
\begin{figure}[t]
    \centering
    \includegraphics[width=\columnwidth]{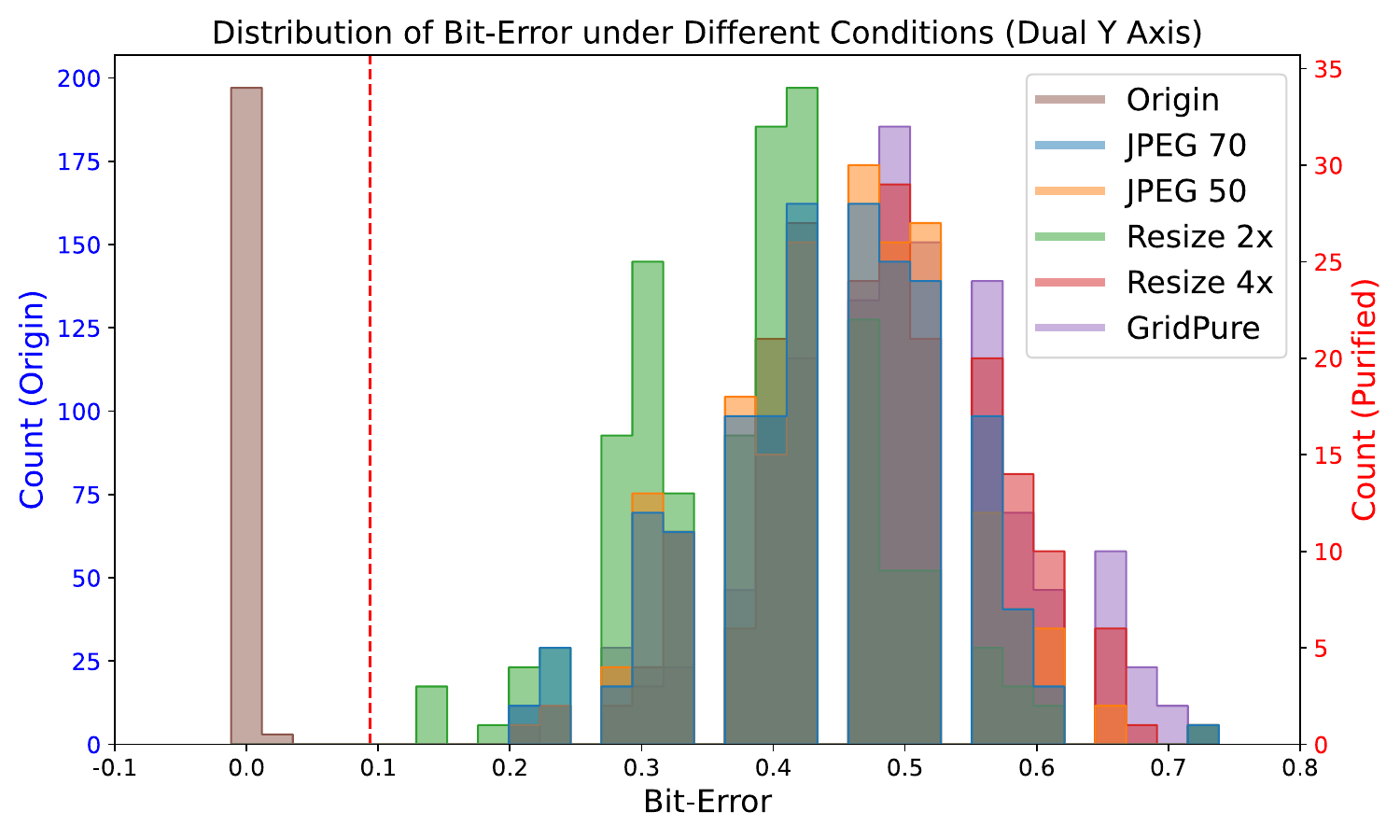}  
    \caption{The distribution of bit-error under different purification settings. ``Origin'' denotes no purification applied. The red dashed line illustrates the PSR bit-error threshold.}
    \label{fig: threshold selection}
\end{figure}

\begin{figure*}[t]
    \centering
    \begin{subfigure}[b]{\textwidth}
        \centering
        \includegraphics[width=0.9\textwidth]{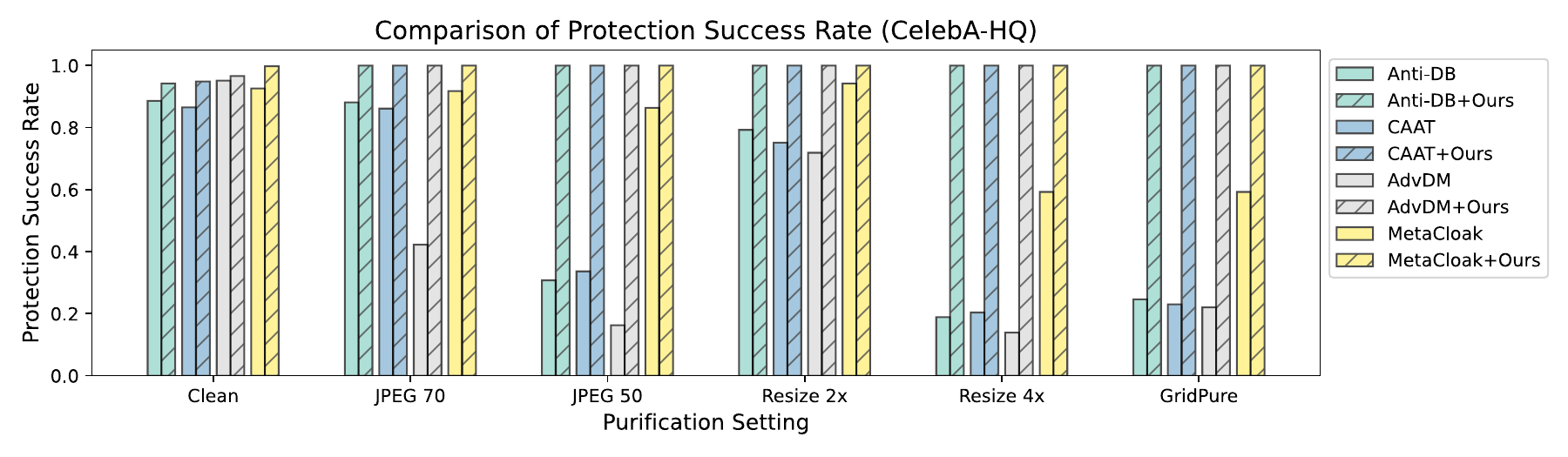}
    \end{subfigure}

    \vspace{-2mm}
    \begin{subfigure}[b]{\textwidth}
        \centering
        \includegraphics[width=0.9\textwidth]{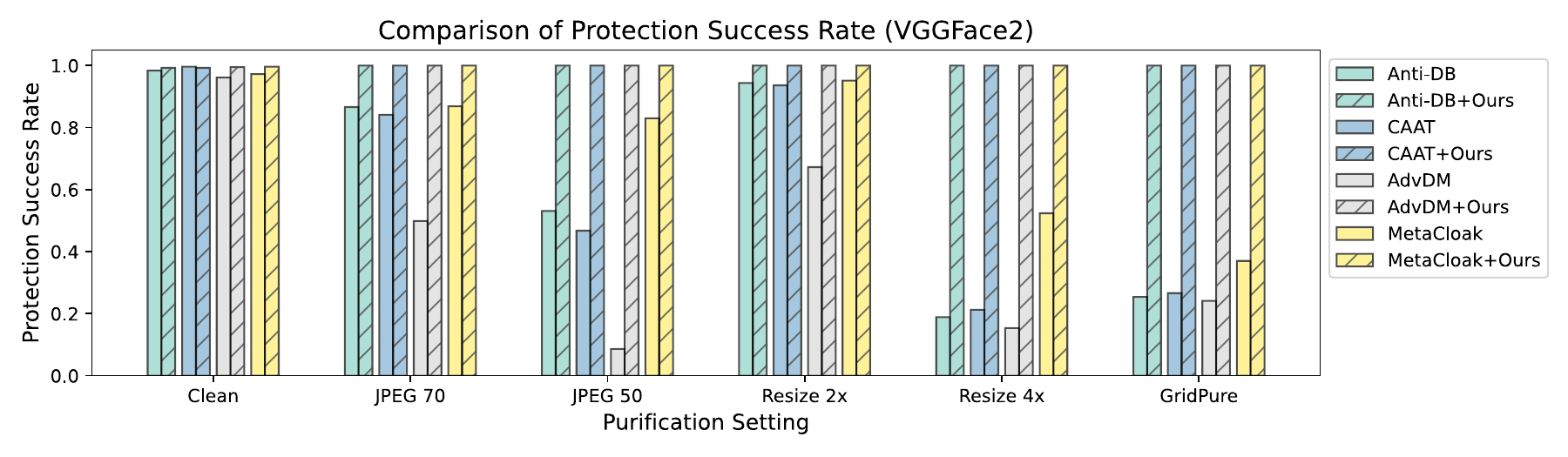}
    \end{subfigure}
     \vspace{-8mm}
    \caption{Comparison of Protection Success Rate for different methods across various purification settings. (Generated by prompt ``a dslr portrait of $sks$person'')}
    \label{fig: attack scenario supp}
\end{figure*}
We adopt Anti-DB for ATP to perturb the images in the CelebA-HQ test set. Then, we adopt purification techniques to purify the image. 
Figure~\ref{fig: threshold selection} shows distinct differences in bit-error rate with and without purification. Since we aim to detect the occurrence of purification through the bit-error threshold, when the occurrence of purification significantly impacts the distribution of bit-errors, setting the threshold becomes a straightforward task.
As a result, \textbf{we set the bit-error threshold of PSR to 3/32}. We adopt this value across different datasets and various protection perturbations, consistently finding that it can be effectively used to reject purification attempts.
\subsection{Frequency-domain Sensitivity Analysis
}\label{sup1_4}

In this section, we analyze why frequency-domain perturbation is inherently more sensitive (i.e., vulnerable) than the pixel-domain perturbation to the purifications.
For pixel-domain purification (e.g., resizing), the vulnerability arises because the Block DCT computes each frequency coefficient as a weighted sum of all pixel values in a block. Thus, even a minor modification to a single pixel can affect all frequency coefficients. For frequency-domain purification (e.g., JPEG), the vulnerability stems from the fact that JPEG compression directly quantizes the frequency coefficients. 
These changes may be smoothed out in the pixel domain due to the inverse DCT and pixel rounding. To support this explanation, we define the change rate as the proportion of frequency coefficients or pixel values that vary before and after purification. 
\begin{table}[t]
\resizebox{\linewidth}{!}{
\centering
\begin{tabular}{l|cc}
\toprule
\textbf{Average Change Rate} & \textbf{Frequency Domain} & \textbf{Pixel Domain} \\
\midrule
4× Resizing          & 0.9714 & 0.7788 \\
JPEG Compression (Q=50) & 0.9594 & 0.8445 \\
\bottomrule
\end{tabular}
}
\caption{The average change rate of coefficients and pixel values after performing different purifications.}
\label{tab:change_rates}
\end{table}
We evaluate it on CelebA-HQ. As shown in Table~\ref{tab:change_rates}, the frequency-domain perturbations have a higher probability of being changed after the purification, resulting in the inherent vulnerability. 

\begin{figure}[tt]
    \centering
    \includegraphics[width=\linewidth]{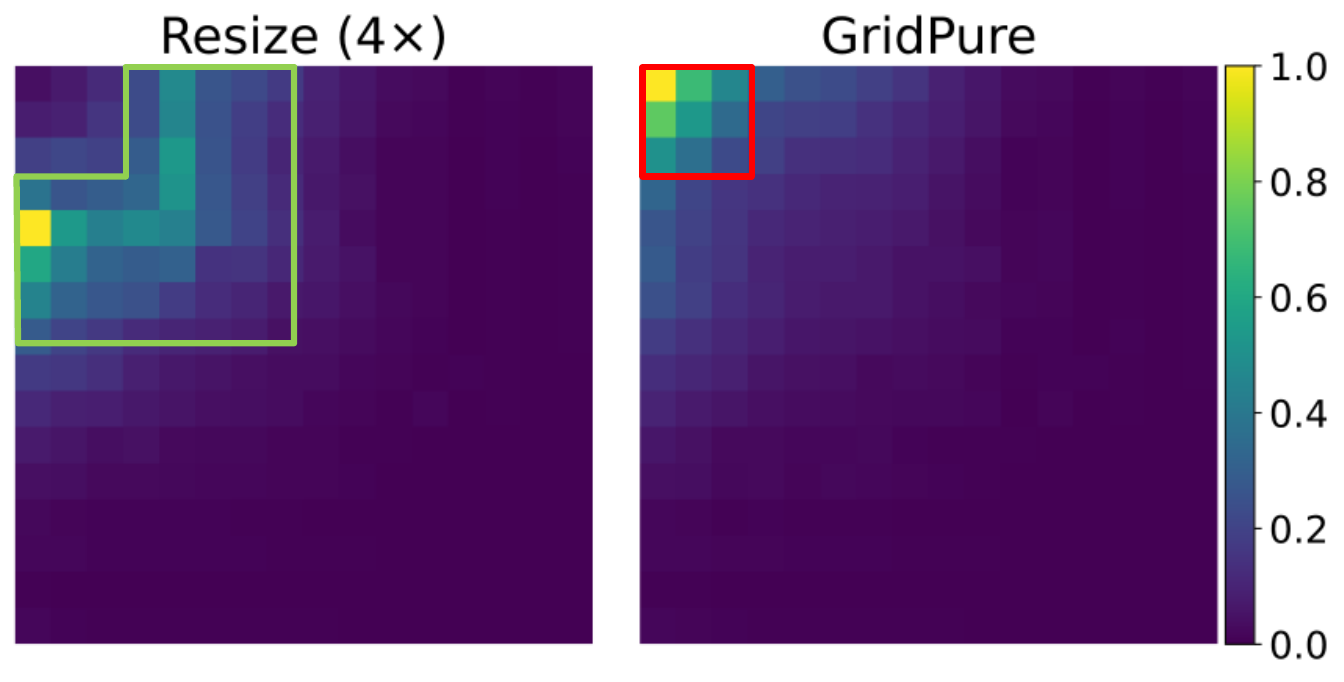}  
    \caption{Visualization of the average normalized variance of the DCT coefficient differences (within 16×16 blocks) before and after purification.}
    \label{fig: spectral}
\end{figure}

\noindent\textbf{Comparison of high- vs. low-frequency resilience to purification.}
While it is commonly assumed that high-frequency components are more vulnerable to traditional purification methods (e.g., resizing), our findings show that advanced purification techniques such as GridPure challenge this assumption.
We want to share that different purification techniques have different preferences for altering frequency bands.  
We computed the average normalized variance of the DCT coefficient differences (within 16×16 blocks) before and after purification. As shown in the Figure~\ref{fig: spectral},  resizing primarily affects 
higher frequency bands (green-box region), whereas GridPure significantly alters low-frequency bands (red-box region).

 Consequently, we adopt a random and uniform perturbation design in this project to ensure sensitivity to different purifications.

\section{More Experiment Results}

\subsection{Influence of Algorithm Design on Mask-guidance}\label{sup2_1}
\begin{figure}[!h]
    \centering
    \begin{subfigure}[b]{0.48\columnwidth}
        \centering
        \includegraphics[width=\textwidth]{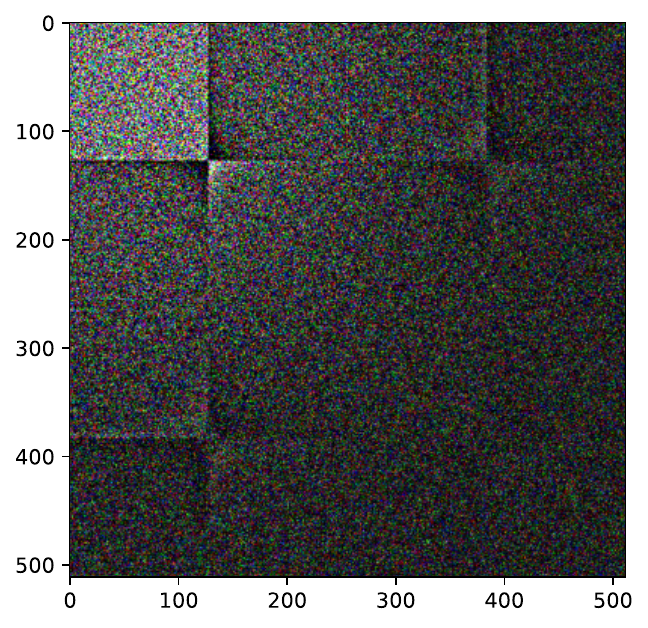}  
        \caption{By Equation 8}
        \label{fig:liqe_comparison}
    \end{subfigure}
    \hspace{0.1mm}
    \begin{subfigure}[b]{0.48\columnwidth}
        \centering
        \includegraphics[width=\textwidth]{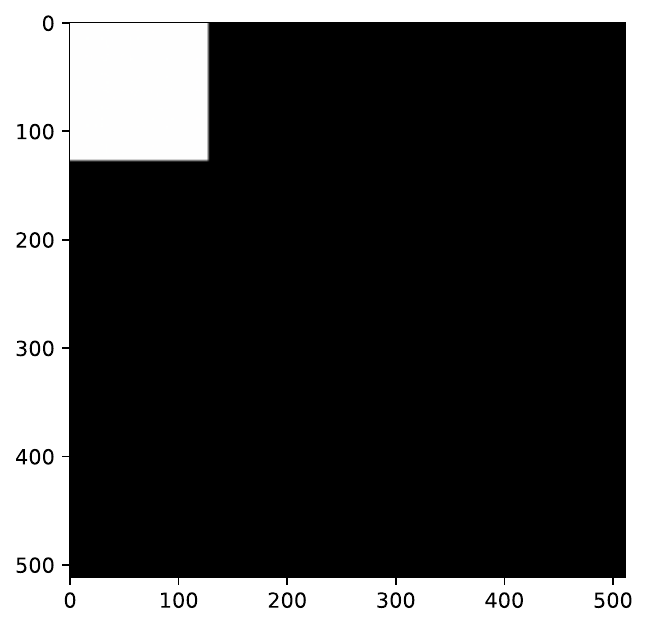}  
        \caption{By Algorithm 1}
        \label{fig:clip_iqac_comparison}
    \end{subfigure}
    \caption{Visualization of change in the frequency domain after the gradient descent. The visualizations depict the changes in frequency domain coefficients after the updates, where black represents no change, and brighter values indicate greater changes.}
    \label{fig: valid mask}
\end{figure}

In this experiment, we aim to demonstrate that Algorithm 1 validates the mask guidance. As outlined in the methodology section, the design of the projected gradient descent algorithm using Equation 8 is intended to invalidate the mask guidance. We verify this through a simulation experiment.

Specifically, we generate random gradients in the frequency domain, ensuring they are concentrated in the top-left 128 × 128 region. A 512 × 512 image is transformed into the frequency domain via DCT, and a one-step gradient descent is conducted using Equation 8 and Algorithm 1 (the step size is 1, and the PGD radius is 1). Subsequently, we visualize the changes in frequency domain coefficients after the single gradient descent step. As illustrated in Figure ~\ref{fig: valid mask}, our algorithm successfully confines the coefficient updates to the designated region in the frequency domain, whereas the original algorithm fails to achieve such precise localization.

\begin{figure}[t]
    \centering
    \includegraphics[width=\linewidth]{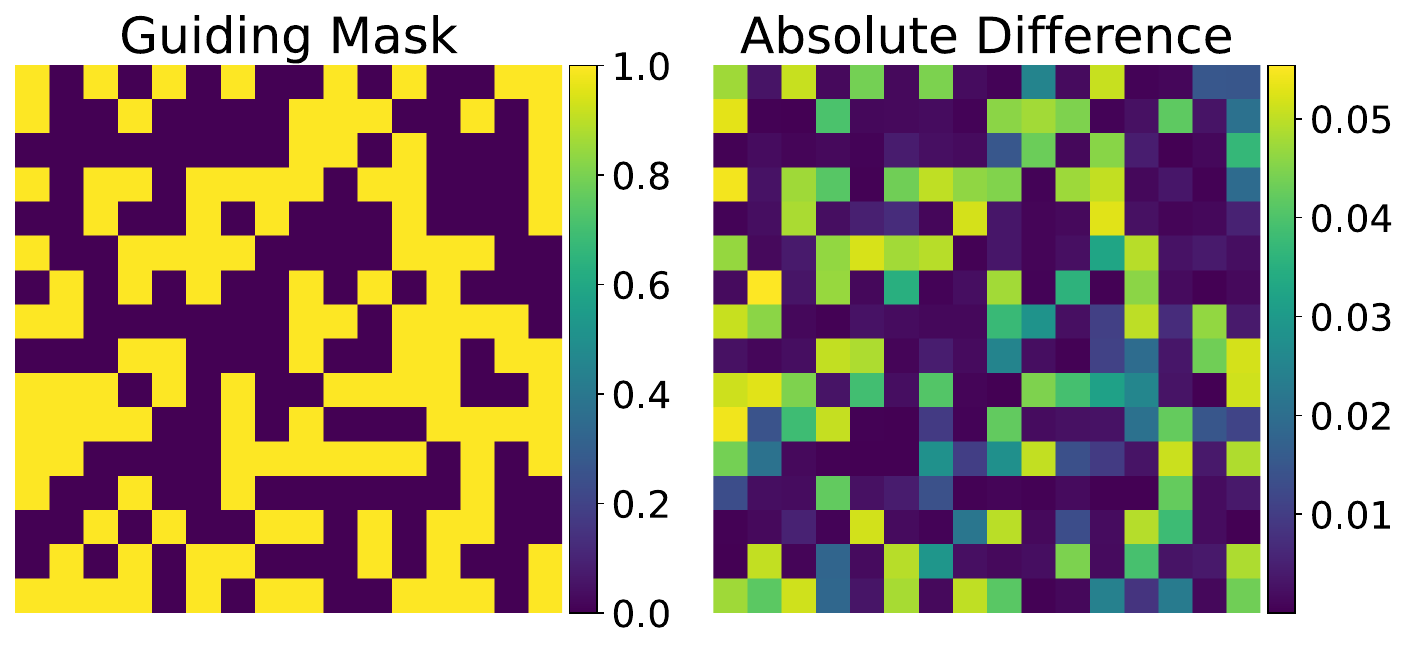}  
    \caption{Visualization of the absolute difference in one 16×16 DCT coefficient map before and after applying the protection perturbation, along with the corresponding guiding mask.}
    \label{fig: vis}
\end{figure}

In addition to the simulation experiment, we also provide a visualization of the absolute difference in one 16×16 DCT coefficient map before and after applying the protection perturbation using Algorithm 1, along with the corresponding guiding mask used during optimization. As illustrated in Figure~\ref{fig: vis}, the perturbation primarily affects regions where the guiding mask is activated (i.e., mask value = 1), confirming that the mask guidance effectively constrains the perturbation by Algorithm 1 (Improved Frequency Domain PGD).

\subsection{Repeat Main Experiments with different prompt}\label{sup2_2}


\begin{table*}[h!]
\centering
\renewcommand{\arraystretch}{1} 
\scalebox{0.8}{
\begin{tabular}{c|cccc|cccc}
\hline

\multirow{2}{*}{} & \multicolumn{4}{c|}{\textbf{CelebA-HQ}} & \multicolumn{4}{c}{\textbf{VGGFace2}} \\ \cline{2-9} 
                  & CLIP-IQAC $\downarrow$ & LIQE $\downarrow$ & ISM $\downarrow$ & FDFR $\uparrow$ & CLIP-IQAC $\downarrow$ & LIQE $\downarrow$ & ISM $\downarrow$ & FDFR $\uparrow$ \\ \hline
AntiDB           & -0.2047	&1.3403	&0.3944	&0.3775	&-0.4274	&1.0228	&0.3233	&0.7950            \\ \hline
\rowcolor{gray!40}
AntiDB+Ours   \rowcolors{green!70}   &-0.3085	&1.1027	&0.3509	&0.4513	&-0.4635	&1.0250	&0.3073	&0.6850            \\ \hline
AdvDM           &-0.2979	&1.0450	&0.3193	&0.6325	&-0.3763	&1.0305	&0.3650	&0.6213\\ \hline
\rowcolor{gray!40}
AdvDM+Ours       &-0.3367	&1.0459	&0.3634	&0.4638	&-0.4703	&1.0126	&0.3103	&0.6538            \\ \hline
CAAT            &-0.1927	&1.3018	&0.4139	&0.3025	&-0.4890	&1.0080	&0.2819	&0.7888  \\ \hline
\rowcolor{gray!40}
CAAT+Ours       &-0.3257	&1.0999	&0.3725	&0.4075	&-0.4902	&1.0192	&0.2914	&0.6963             \\ \hline
Metacloak        &-0.2573	&1.4254	&0.3892	&0.5000	&-0.4485	&1.1075	&0.3513	&0.8613  \\ \hline
\rowcolor{gray!40}
Metacloak+Ours   &-0.4049	&1.0891	&0.3488	&0.7975	&-0.4694	&1.0447	&0.3500	&0.8875               \\ \hline
\end{tabular}
}
\caption{Quantitative results for CelebA-HQ and VGGFace2 datasets across various metrics. (Generated by prompt ``a dslr portrait of $sks$ person'')}
\label{tab: MR_sup}
\end{table*}

Following the experimental setup described in~\cite{ANTIDB, METACLOAK}, we evaluate the protection performance of our method using an alternative prompt: ``a dslr portrait of $sks$ person", to generate individual images. We adopt the same experimental setup described in Section 4, with the sole distinction being the prompt utilized.

Figure~\ref{fig: attack scenario supp} shows that, with the new prompt, ATP continues to safeguard individual image generation effectively. This is because the integrity-check mechanism prevents generation before the prompt is utilized, ensuring that the performance of this mechanism remains unaffected by variations in the prompt.
Table~\ref{tab: MR_sup} reveals that, under the new prompt, ATP still performs comparably to the original protection perturbation approaches when the purification techniques are not applied.

\subsection{Generalizability Analysis}\label{sup2_3}

We report the protection performance of ATP (CAAT) trained on SD2.1 when applied to a different diffusion model (SD1.5) and  personalization method (SVDiff~\cite{han2023svdiff}) using CelebA in Table~\ref{tab:single_column_combined}. We compare the protection performance of ATP against that of the unprotected baseline (i.e., without any perturbation applied).

\begin{table}[h]
\scriptsize
\resizebox{\linewidth}{!}{
\centering
\begin{tabular}{l|cccc}
\toprule
\multicolumn{5}{c}{\textbf{SD1.5 + DreamBooth}} \\
\midrule
           & CLIP-IQAC↓ & LIQE↓ & ISM↓ & FDFR↑ \\
Origin      & 0.5007     & 4.5427 & 0.6824 & 0.0125 \\
ATP        & -0.2893    & 1.1929 & 0.4329 & 0.2988 \\
\midrule
\multicolumn{5}{c}{\textbf{SD2.1 + SVDiff}} \\
\midrule
           & CLIP-IQAC↓ & LIQE↓ & ISM↓ & FDFR↑ \\
Origin      & 0.3837     & 4.3704 & 0.6679 & 0.1338 \\
ATP        & -0.3307    & 1.0484 & 0.3861 & 0.5575 \\
\bottomrule
\end{tabular}}
\caption{Protection Performance of ATP when generation model and algorithm are changed.}
\label{tab:single_column_combined}

\end{table}
 The results demonstrate that ATP is generalizable across diffusion models and personalization techniques.

\subsection{Performance Trade-off on Mask Ratio}\label{sup2_4}

The authorization and protection perturbations in the frequency domain can be distinguished based on the random mask $M$. The mask ratio $p$ controls the region in the frequency domain used for authorization versus protection.  This experiment shows that adjusting the mask ratio achieves a performance balance between protection and authorization for the ATP.

For example, as the mask ratio increases, a larger portion of the frequency domain will be allocated to authorization. As shown in Table~\ref{maskratio} and Table~\ref{sup_maskratio}, the increase in mask ratio leads to a decrease in bit-error, reflecting an improvement in message embedding accuracy. It also decreases protection performance, as LIQE, CLIP-IQAC, ISM, and FDFR scores indicate. 
Thus, we adopt a mask ratio of 0.5 as the default setting to achieve a balanced trade-off between authorization and protection performance.
\subsection{Performance Trade-off on Block Size}\label{sup2_5}

\begin{figure}[!t]
    \centering
    \includegraphics[width=\linewidth]{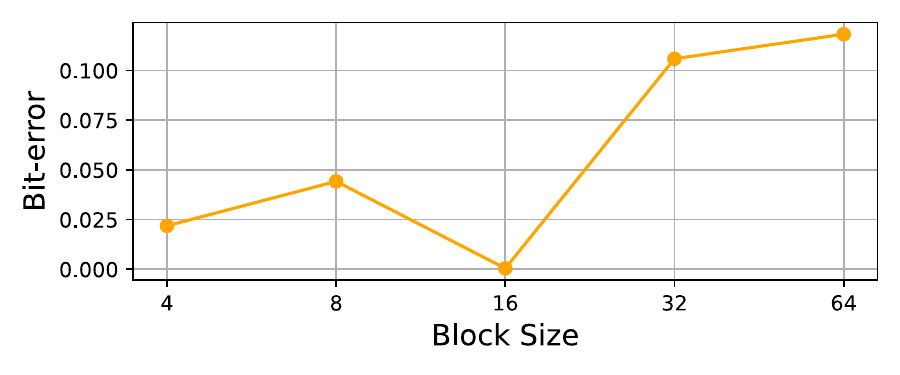}  
    \caption{The Bit-error variation under different block size.}
    \label{fig:blockvar}
\end{figure}

The frequency domain transformation is achieved by BDCT. One of the hyperparameters for it is the size of the Block. In this section, we report the influence of this hyperparameter on the information hiding of authorization perturbation.
We train the authorization model using different block sizes and evaluate it on CelebA-HQ.
Figure~\ref{fig:blockvar} visualizes the variation in Bit-error under different block sizes. Since the block is square-shaped, we use its side length to represent the block size. It can be observed that a size of 16$\times$16 yields the lowest Bit-error, supporting our design choice adopted in the project.

\subsection{Protection Performance Achieved Using Only Authorization Perturbation}\label{sup2_6}

In this section, we discuss the protection performance when we don't include protection perturbation in the ATP design.
We compare the protection performance of images with no perturbation, images with authoirzation perturbation and images with ATP (taking CAAT as protection perturbaiton) in CelebA-HQ.
As shown in the Table~\ref{tab: alone}, authorization perturbation alone fails to provide strong protection when purification is not applied.

\begin{table}[h]
\resizebox{\linewidth}{!}{
\centering
\small
\begin{tabular}{lcccc}
\toprule
 & \textbf{CLIP-IQAC$\downarrow$} & \textbf{LIQE$\downarrow$} & \textbf{ISM$\downarrow$} & \textbf{FDFR$\uparrow$} \\
\midrule

Origin (No Perturb)  &  0.4659 & 4.2340 & 0.7053 & 0.0975 \\
Authorization Alone       &  0.3258 & 3.6935 & 0.6414 & 0.1075 \\
ATP\ (CAAT)           & -0.3568 & 1.0768 & 0.4315 & 0.6338 \\
\bottomrule
\end{tabular}}
\caption{The protection performance using only the authorization perturbation is significantly worse than that of ATP.}
\label{tab: alone}
\end{table}

As a result, the combination of protection perturbation and authorization perturbation (ATP) is crucial for achieving reliable protection.

\subsection{Repeat Experiments on VGGFace2}\label{sup2_7}

\begin{figure}[tbp]
    \centering
    \includegraphics[width=0.9\columnwidth]{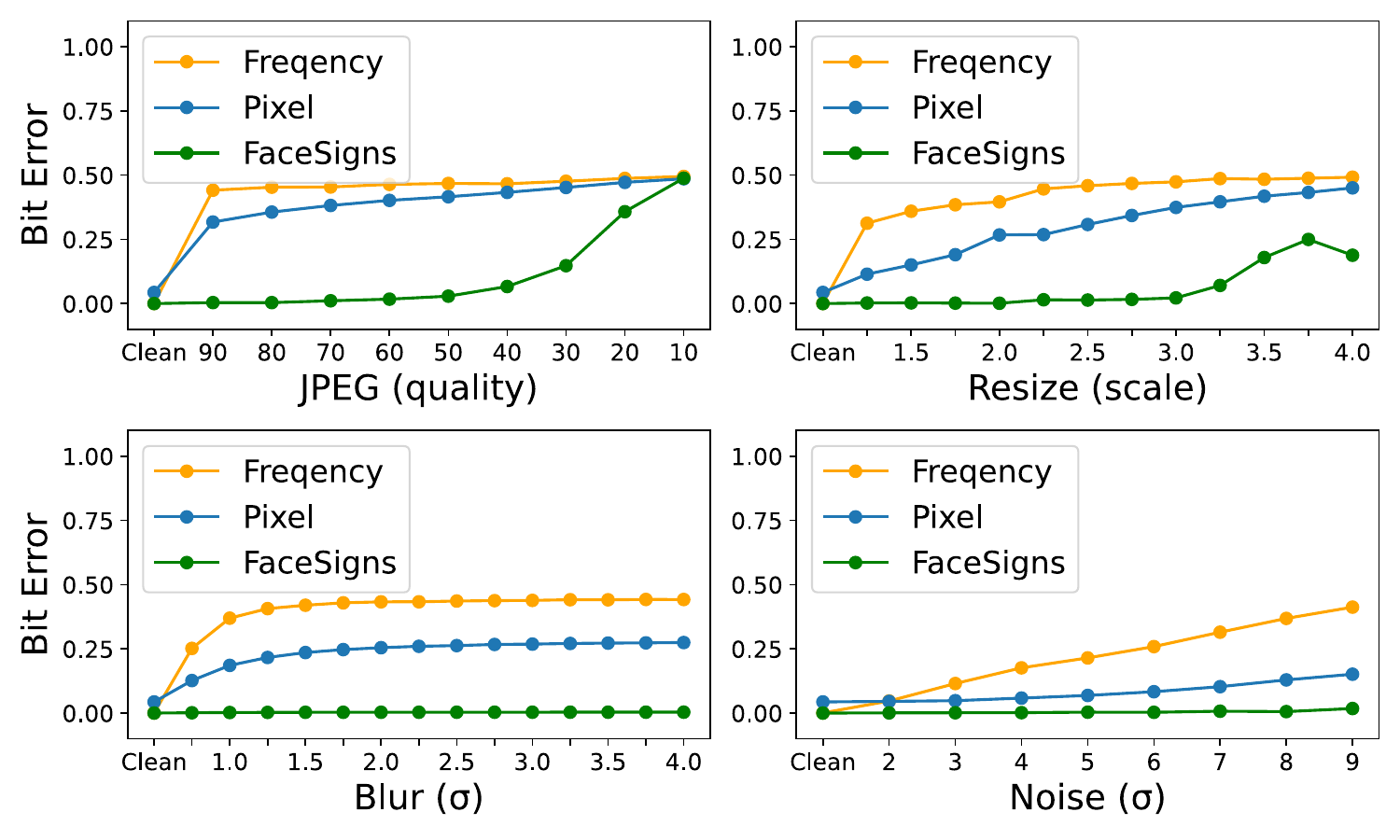}  
    \caption{Sensitivity of ATP to different types of purification. 
    The x-axis indicates the hyperparameter of different purifications, while the y-axis indicates the Bit-error. The images are from VGGFace2.}
    \label{fig:Ab_pixelfreq_sup}
\end{figure}

We repeat the experiment on VGGFace2 to further validate the credibility of our conclusions in Section 4. We adopt the same experimental setup described in Section 4, with the sole distinction being the dataset utilized. 
The experiment results are shown in Table~\ref{tab: sup_MC_2} and Figure~\ref{fig:Ab_pixelfreq_sup}.
\begin{table}[!tbp]
    \centering
 
    \resizebox{\linewidth}{!}{
    \begin{tabular}{lccccccc}
        \toprule[1.5pt]
        \textbf{Ratio} & \textbf{Bit-error ($e^{-3}$) $\downarrow$} & \textbf{CLIP-IQAC $\downarrow$}  & \textbf{LIQE $\downarrow$} &\textbf{ISM $\downarrow$} & \textbf{FDFR $\uparrow$}  \\
        \midrule
        0.25   &0.7813	&-0.3561	&1.0471	&0.3805	&0.6400 \\
        0.50   &0.4688	&-0.3139	&1.0741	&0.4647	&0.5213 \\
        0.75   &0.3125	&-0.1480	&1.3582	&0.5765	&0.2225 \\
        \bottomrule[1.5pt]
    \end{tabular}}
    \caption{Performance comparison for different mask ratios on CelebA-HQ.}
    \label{maskratio}
\end{table}
\begin{table}[!tbp]
    \centering
 
    \resizebox{\linewidth}{!}{
    \begin{tabular}{lccccccc}
        \toprule[1.5pt]
        \textbf{Ratio} & \textbf{Bit-error ($e^{-3}$) $\downarrow$} & \textbf{CLIP-IQAC $\downarrow$}  & \textbf{LIQE $\downarrow$} &\textbf{ISM $\downarrow$} & \textbf{FDFR $\uparrow$}  \\
        \midrule
        0.25   &3.1250	&-0.422682	&1.135657	&0.205465	&0.91375 \\
        0.50   &0.7813	&-0.386397	&1.098367	&0.254911	&0.81875 \\
        0.75   &1.2500	&-0.371436	&1.03989	&0.340458	&0.70625 \\
        \bottomrule[1.5pt]
    \end{tabular}}
    \caption{Performance comparison for different mask ratios on VGGFace2.}
    \label{sup_maskratio}
\end{table}
\begin{table}[!tbp]
    \centering
    \renewcommand{\arraystretch}{0.9} 
    \resizebox{0.7\linewidth}{!}{
    \begin{tabular}{c|ccc|c}
        \toprule[1.5pt]
        & BDCT & Improved-PGD & Mask & Bit-error ($e^{-3}$)$\downarrow$  \\
        
        \hline
        (a) & & & & 366.09  \\
        \hline
        (b) & & & $\checkmark$ & 43.594  \\
        \hline
        (c) & $\checkmark$ & & & 503.75  \\
        \hline
        (d) & $\checkmark$ & & $\checkmark$ & 79.844  \\
        \hline
        Ours & $\checkmark$ & $\checkmark$ & $\checkmark$ & 0.7813  \\
        \bottomrule[1.5pt]
    \end{tabular}}
    \caption{Comparison of different fusion designs with Bit-error values on VGGFace2. }
    \label{tab: sup_MC_2}
\end{table}

\subsection{Visualization of Perturbed Images}\label{sup2_8}

\begin{figure*}[ht]
    \centering
    \includegraphics[width=0.75\linewidth]{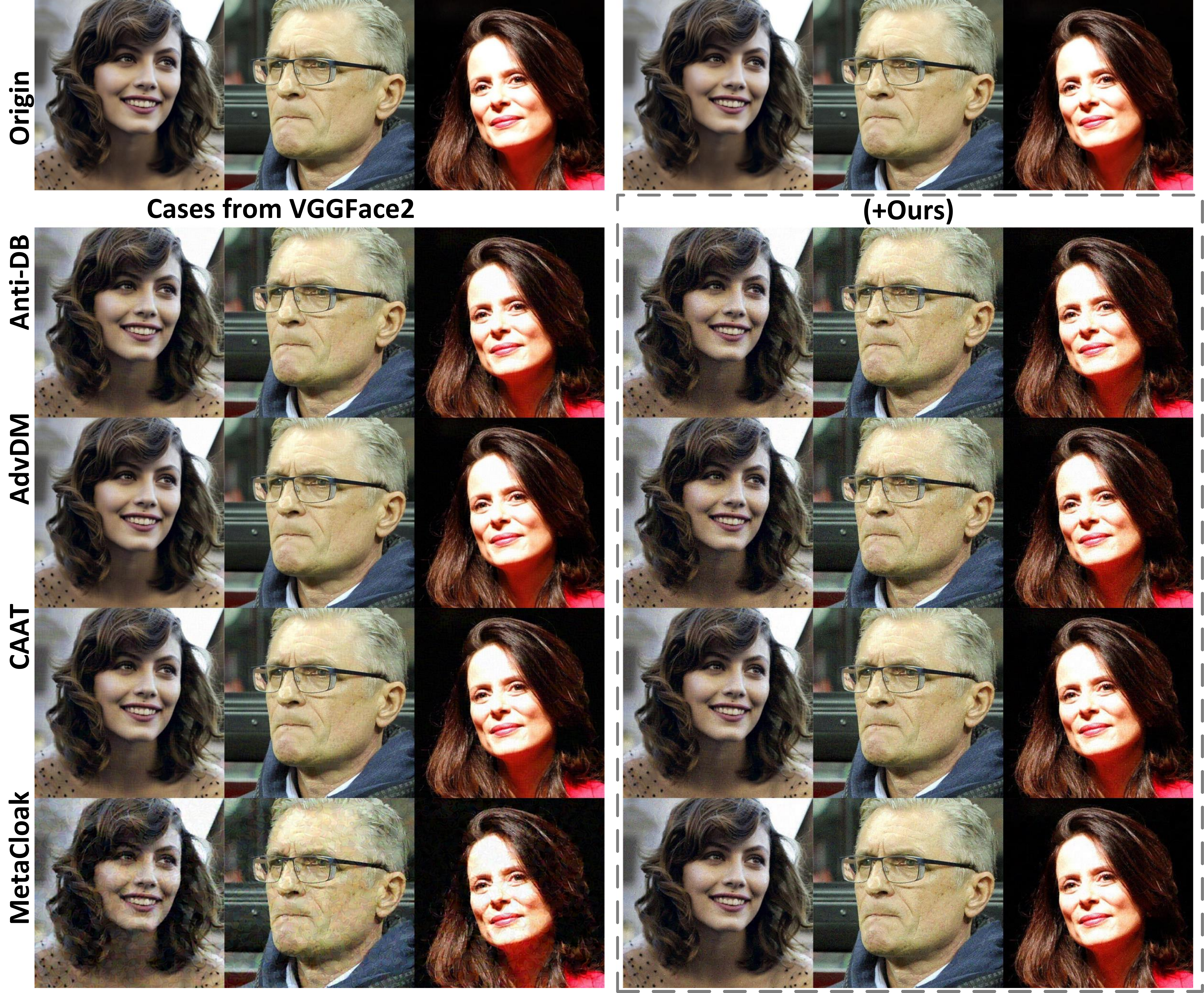}  
    \caption{Perturbed images of different methods from two datasets.}
    \label{fig: perturbed imgs}
\end{figure*}

In Figure~\ref{fig: perturbed imgs}, we present perturbed images generated using different methods from the CelebA-HQ and VGGFace2 datasets. We observe that while perturbations are difficult to detect at normal scales, they become noticeable when viewed at an enlarged scale. This remains an unresolved challenge in the field and a focus of our future research efforts.
\subsection{Visualization of Generation Results Applied Purification Techniques}\label{sup2_9}

\begin{figure*}[ht]
    \centering
    \includegraphics[width=0.8\linewidth]{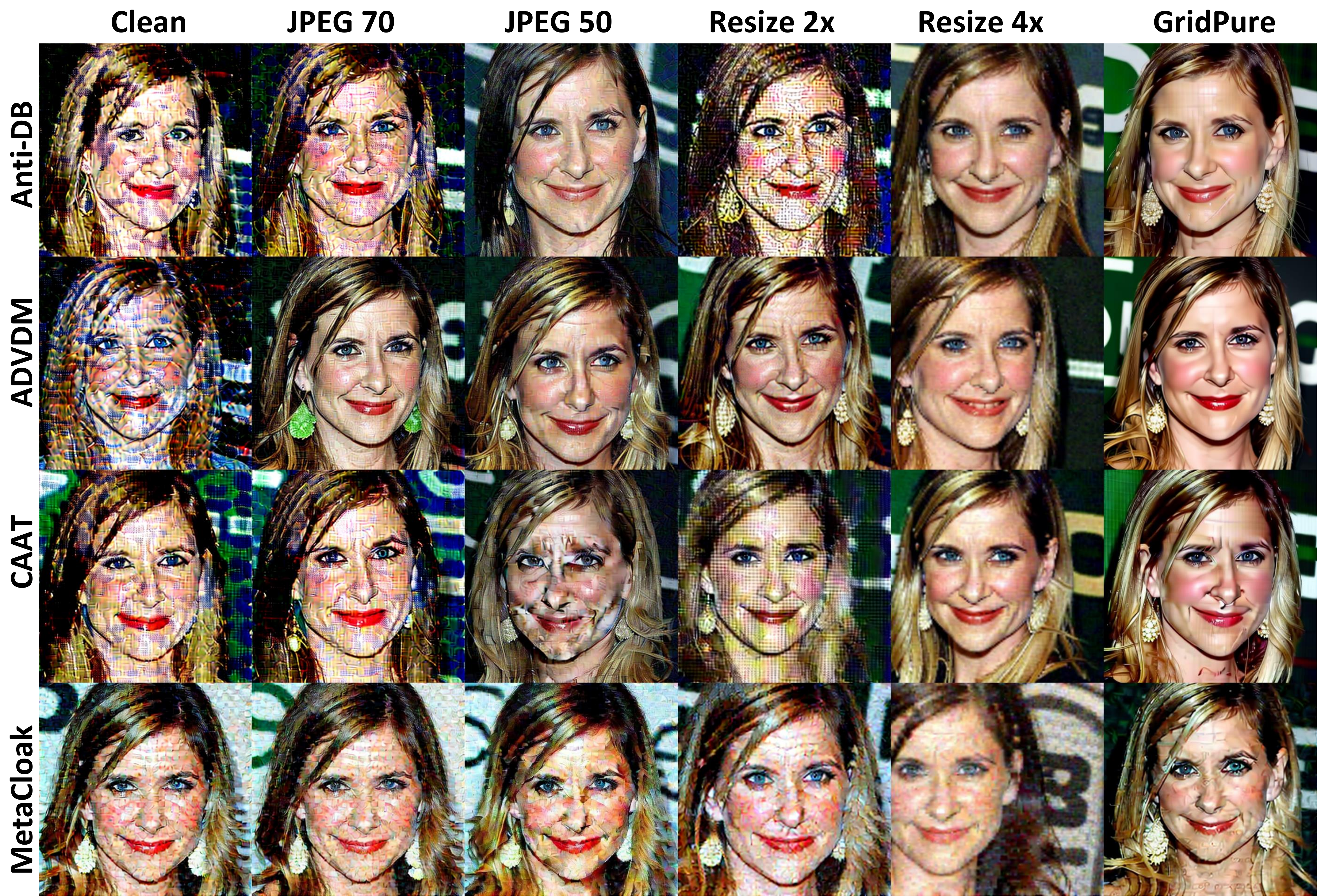}  
    \caption{Visual cases showing the purification results bypassing the protection mechanisms on images from the CelebA-HQ dataset. ``Clean'' indicates no purification applied.}
    \label{fig: purified_vis_celeba}
\end{figure*}

\begin{figure*}[ht]
    \centering
    \includegraphics[width=0.8\linewidth]{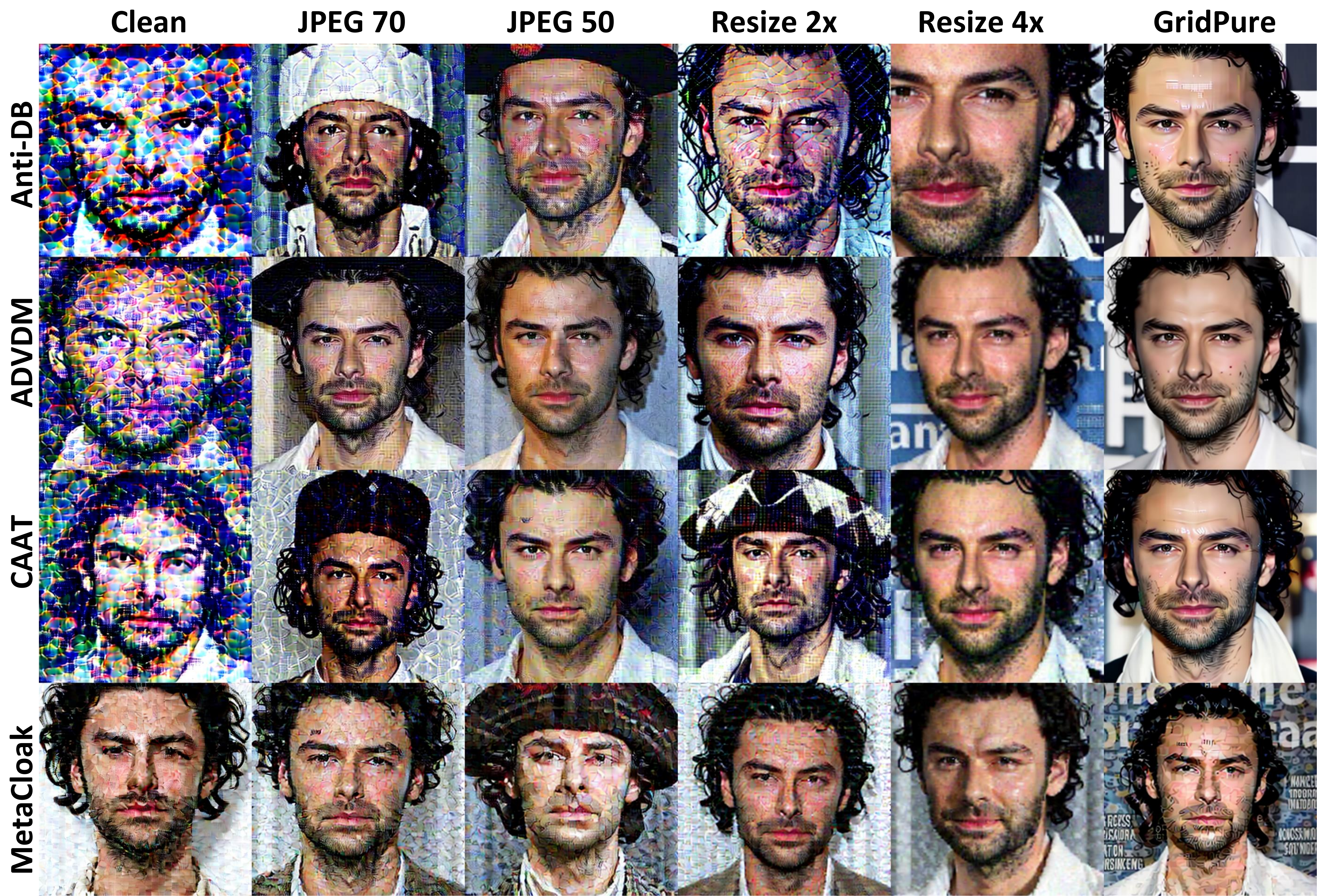}  
    \caption{Visual cases showing the purification results bypassing the protection mechanisms on images from the VGGFace2 dataset. ``Clean'' indicates no purification applied.}
    \label{fig: purified_vis_vgg}
\end{figure*}

We prepared visual cases to illustrate how purification techniques bypass existing protection mechanisms. Specifically, we present the results of individual image generation for images from CelebA-HQ and VGGFace2 after applying different protection perturbation algorithms. As demonstrated in Figure~\ref{fig: purified_vis_celeba} and Figure~\ref{fig: purified_vis_vgg}, purification can bypass the protection provided by protection perturbation, compromising the safeguarding of individual image generation.

\subsection{More Qualitative Results of Main Experiments}\label{sup2_10}

\begin{figure*}[ht]
    \centering
    \includegraphics[width=0.85\linewidth]{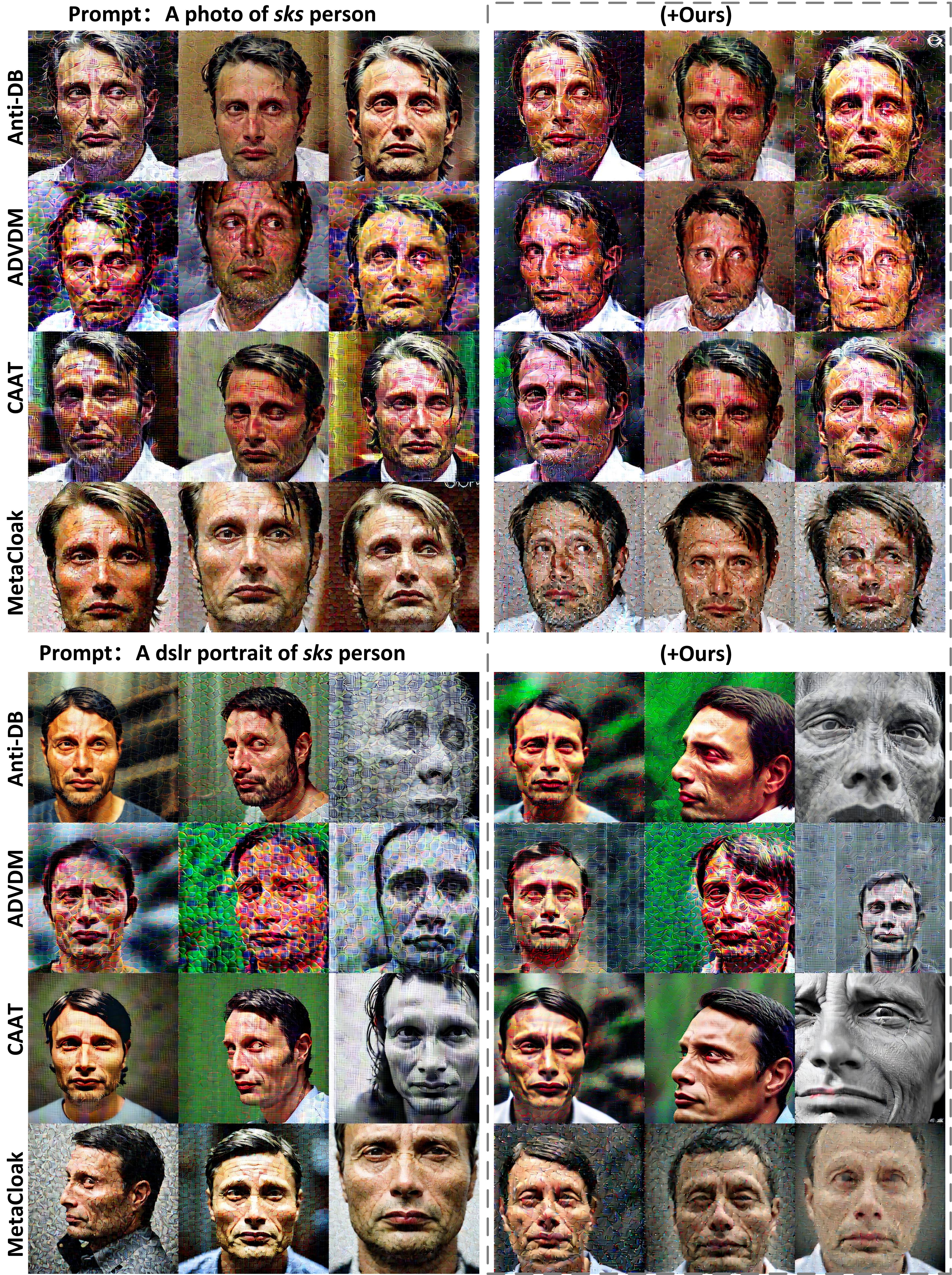}  
    \caption{Qualitative comparison of original perturbation algorithms and their ATP modified versions in CelebA-HQ.}
    \label{fig: sup_celeba_1}
\end{figure*}

\begin{figure*}[ht]
    \centering
    \includegraphics[width=0.85\linewidth]{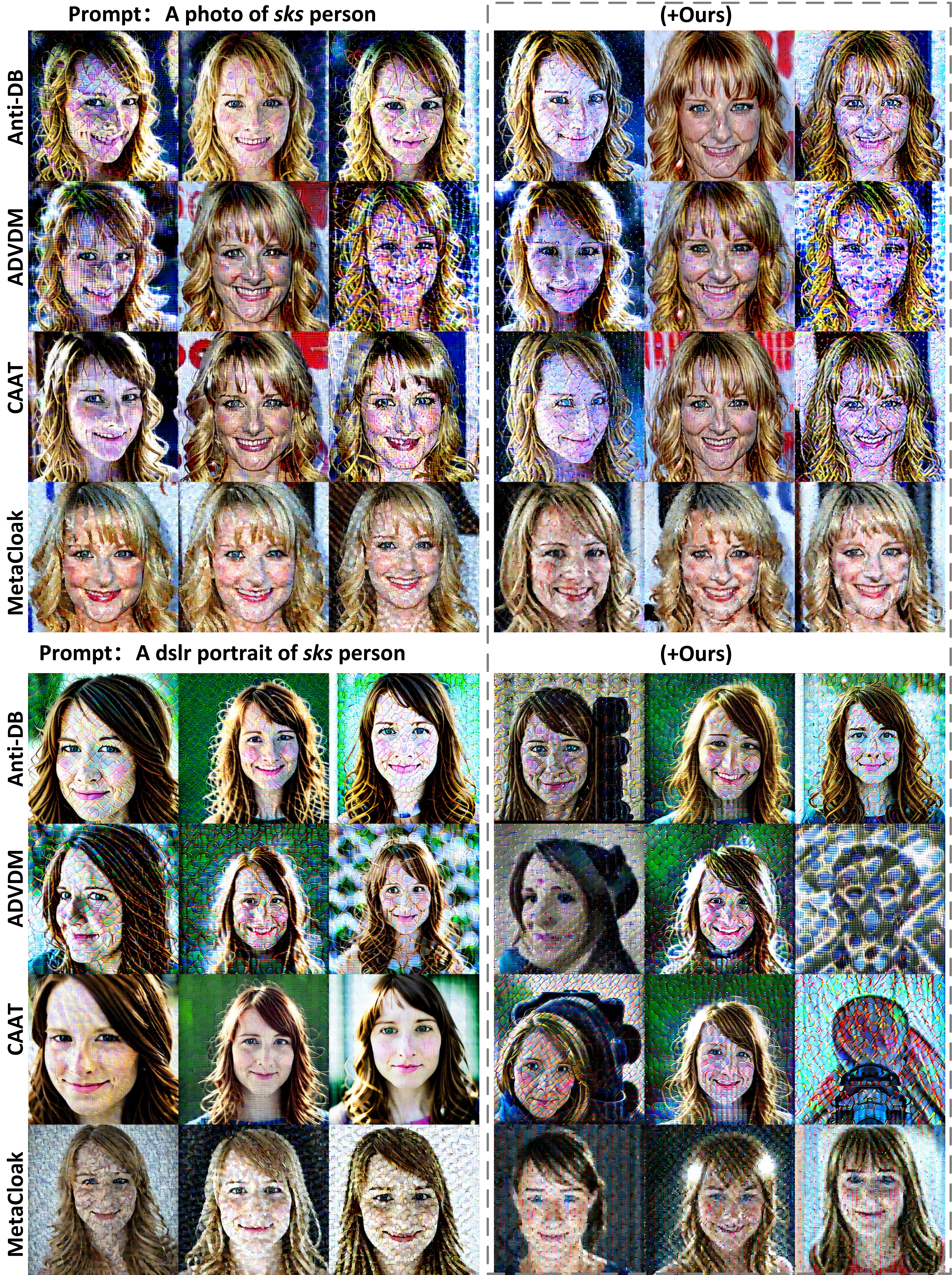}  
    \caption{Qualitative comparison of original perturbation algorithms and their ATP modified versions in CelebA-HQ.}
    \label{fig: sup_celeba_2}
\end{figure*}

\begin{figure*}[ht]
    \centering
    \includegraphics[width=0.85\linewidth]{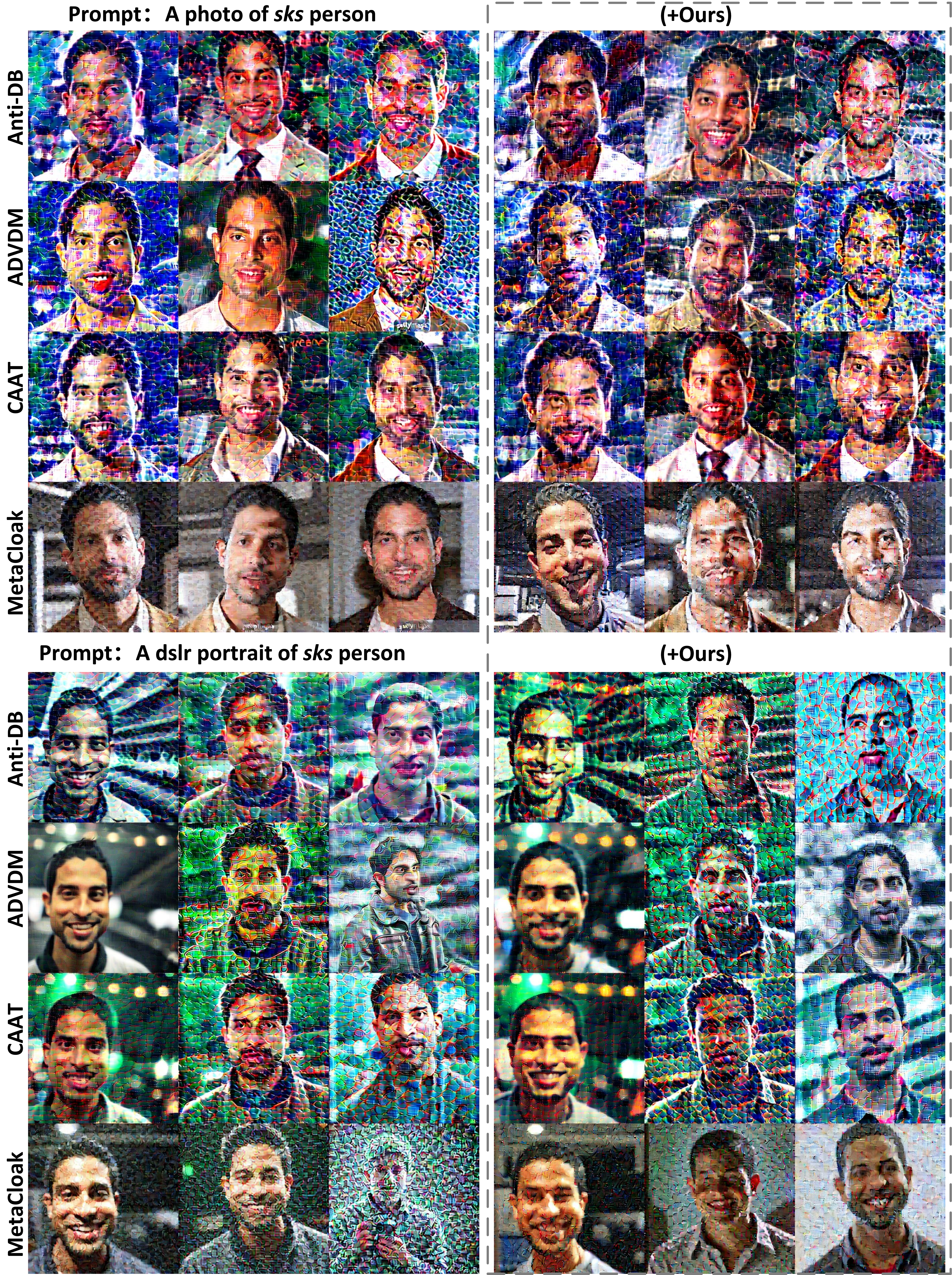}  
    \caption{Qualitative comparison of original perturbation algorithms and their ATP modified versions in VGGFace2}
    \label{fig: sup_vgg_1}
\end{figure*}

\begin{figure*}[ht]
    \centering
    \includegraphics[width=0.85\linewidth]{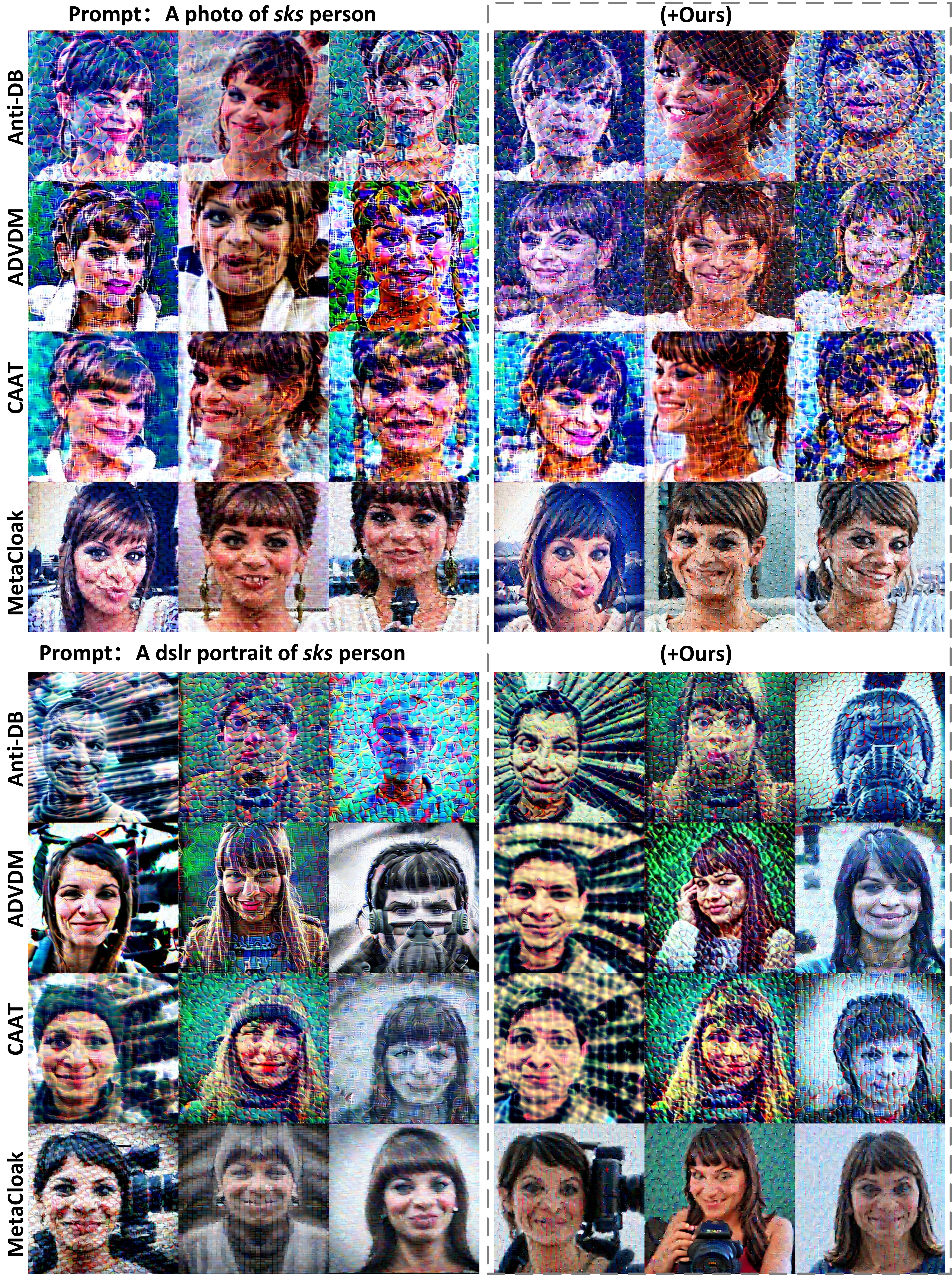}  
    \caption{Qualitative comparison of original perturbation algorithms and their ATP modified versions in VGGFace2}
    \label{fig: sup_vgg_2}
\end{figure*}

We present additional qualitative comparison results across various methods under two datasets (i.e., CelebA-HQ, VGGFace2) and two different prompts (i.e., a photo of $sks$ person, a dslr portrait of $sks$ person) in Figure~\ref{fig: sup_celeba_1}, Figure~\ref{fig: sup_celeba_2}, Figure~\ref{fig: sup_vgg_1}, and Figure~\ref{fig: sup_vgg_2}.

\subsection{Scalability and Computational Efficiency Analysis}\label{sup2_11}

\noindent\textbf{Scalability.} A safety checker is deployed by the widely used diffusion model library ``diffusers", which takes up 1159.60 MB.  The authorization model only takes up 1.58 MB, which should be affordable by the service providers.

\noindent\textbf{Computational Efficiency.}
The ATP requires extra time in authorization message hiding and verification.
With batch size = 4, the averaged inference time costs are: Autoencoder encoding/decoding: 0.0201s/0.0274s; BDCT + IBDCT: 0.0016s.
In the protection phase, ATP using CAAT as protection perturbation performs autoencoder encoding once and applies mask-guided PGD, which requires two additional BDCT+IBDCT operations per PGD step. This results in \textbf{0.38\% increase} of the total protection time compared to the original CAAT protection (77.33s). In the generation phase, autoencoder decoding is performed once. When considering a generation method like DreamBooth (341.9s), the added decoding introduces \textbf{0.008\% increase}.

\end{document}